\documentclass[amsmath,amssymb,showkeys,superscriptaddress,floatfix,prd,10pt]{revtex4}
\usepackage{graphicx,epstopdf}
\usepackage{subfig}
\usepackage{xcolor}
\usepackage[colorlinks=true, urlcolor=blue, linkcolor=blue, citecolor=green]{hyperref}
\def\pslash{p\!\!\!\slash }
\def\qslash{q\!\!\!\slash }
\def\xslash{x\!\!\!\slash }
\def\eslash{\varepsilon\!\!\!\slash }

\def\vel{\left|}
\def\ver{\right|}

\begin{document}

\title{The electromagnetic multipole moments of the possible charm-strange pentaquarks in light-cone QCD}
\author{K. Azizi}%
\email[]{kazizi@dogus.edu.tr}
%%%%%
\affiliation{Department of Physics, Dogus University, Acibadem-Kadikoy, 34722 
Istanbul, Turkey}
\affiliation{School of Physics, Institute for Research in Fundamental Sciences (IPM),
P.~O.~Box 19395-5531, Tehran, Iran}
\author{U. \"{O}zdem}%
\email[]{uozdem@dogus.edu.tr}
\affiliation{Department of Physics, Dogus University, Acibadem-Kadikoy, 34722 
Istanbul, Turkey}

\date{\today}
 
\begin{abstract}
We investigate the electromagnetic properties of possible charm-strange 
pentaquarks in the framework of the light-cone QCD sum rule using the photon distribution amplitudes. 
In particular, by calculating the corresponding electromagnetic form factors defining the radiative 
transitions under consideration we estimate the magnetic dipole and electric quadrupole moments of the
pentaquark systems of a charm, an anti-strange and three light quarks. We observe that the values of 
magnetic dipole moments are considerably large, however, the quadrupole moments are very small. 
Any future measurements of the electromagnetic parameters under consideration and 
comparison of the obtained data with the theoretical predictions can shed light on 
the quark-gluon organization as well as the nature of the pentaquarks.
\end{abstract}

\maketitle

\section{Introduction}
Although the existence of the exotic states was predicted  many decades ago 
by Jaffe~\cite{Jaffe:1976ih}, this subject 
has experienced two revolutions in the last two decades. The first one was the discovery of 
the famous X(3872) tetraquark state by Belle 
experiment~\cite{Choi:2003ue} in 2003. The second revolution was in 2015 when
the LHCb Collaboration announced the observation of the hidden-charmed
$P_c^+(4380)$ and $P_c^+(4450)$ pentaquarks with the spin-parities 
$J^P =\frac{3}{2}^-$ and $\frac{5}{2}^+$, 
respectively~\cite{Aaij:2015tga}.
Now we have many exotic states discovered via different experiments.
For more information on the experimental and 
 theoretical progresses on the features of these new particles see for instance Refs.~\cite{Faccini:2012pj,Esposito:2014rxa,Chen:2016qju,Ali:2017jda,
Esposito:2016noz,Olsen:2017bmm,Lebed:2016hpi,Guo:2017jvc,Nielsen:2009uh,Swanson:2006st,
 Voloshin:2007dx,Klempt:2007cp,Godfrey:2008nc}.
Despite a lot of the experimental and theoretical efforts,
since the discovery of the first exotic state in 2003, 
on the physical properties of the non-conventional or exotic
states, their internal quark-gluon organization, nature 
and quantum numbers are not well-established and there are many
questions to be answered. The spectroscopic parameters of these states
have been widely investigated both in theory and experiment.
Many suggestions on the internal quark structure of the exotic 
states give consistent mass results with the experimental data.
This prevents us to have exact assignments on the internal 
structure, nature and quantum numbers of the exotic states~
\cite{Azizi:2018dva,Azizi:2018bdv,Azizi:2017bgs,Azizi:2016dhy}. 
Hence, we need move investigations on the fundamental
interactions of these states with each other and other known particles.
Among these interactions are the electromagnetic interactions 
of these states and their radiative decays.
Analysis of the electromagnetic and multipole moments of the 
exotic states can help us get valuable knowledge about the 
electromagnetic properties of these states, the charge distributions
inside them, their charge radius and 
geometric shapes and finally their internal substructure.

As we mentioned above, the electromagnetic multipole moments are straight-forwardly connected
with the charge and current distributions in the particles 
and these observables contain important information on the internal spatial quarks and gluons  distributions of the particles.
Their sign and magnitude encode valuable information on shape, structure and  size of hadrons. 
There exist a lot of  studies in the literature in which the electromagnetic properties of conventional 
hadrons are studied and electromagnetic multipole moments are obtained,
but unluckily our knowledge on the electromagnetic multipole moments of the non-conventional hadrons are very limited.
There exist only few studies in the literature devoted to the study of  the electromagnetic multipole moments
of the exotic states~\cite{Kim:2003ay,Ozdem:2017exj,Wang:2017dce,Azizi:2018mte,Agamaliev:2016wtt,
Huang:2003bu,Liu:2003ab,Wang:2005gv,Ozdem:2018qeh,Wang:2005jea,Wang:2016dzu,
Bijker:2004gr,Li:2003cb,Ozdem:2017jqh}.
%%%%%%%%%%%%%%%%%
Theoretical works can play important roles in this respect since direct experimental information about the electromagnetic 
multipole moments of exotic particles is very limited.
In this study, the electromagnetic multipole moments of the charm-strange
pentaquark states (hereafter we will denote these states as $P_{c\bar s}$) are extracted by 
using the diquark-diquark-antiquark picture 
in the framework of the light-cone QCD sum rule (LCSR)
(for more about this method see, e.g.,~\cite{Chernyak:1990ag, Braun:1988qv, Balitsky:1989ry} 
and references therein).
This method has already been successfully applied to investigate the dynamical and
statical properties of hadrons for many years such as, coupling constants, form factors, 
masses and electromagnetic multipole moments. 
 In the LCSR, the features of the particles under investigations are defined based on  the light-cone distribution 
amplitudes (DAs) that determine the matrix elements of the nonlocal operators
between vacuum and corresponding particle states. Therefore, any uncertainty in these parameters 
affects the predictions on the electromagnetic multipole moments.

The rest part of the paper is coordinated in the following way: In section II, we present the result for
the $P_{c\bar s}$ pentaquarks electromagnetic multipole moments in the LCSR method. 
Section III is devoted to the numerical analysis of the obtained sum rules. Section IV includes
our concluding remarks.  The QCD sum rules of the electromagnetic form factors entering the epressions 
of the magnetic dipole and electric quadrupole moments are collected in the Appendix.

\section{Formalism}
  In order to determine  the electromagnetic multipole moments in the
framework of the LCSR, we take into consideration the following
two-point correlation function:
\begin{equation}
 \label{edmn01}
\Pi _{\mu \nu }(q)=i\int d^{4}xe^{ip\cdot x}\langle 0|\mathcal{T}\{J_{\mu}(x)
\bar J_{\nu }(0)\}|0\rangle_{\gamma}, 
\end{equation}%
where $\gamma$ means the external electromagnetic field,  $J_{\mu}$ is the interpolating current of $P_{c\bar s}$ 
pentaquark with spin-$\frac{3}{2}$. 
In the diquark-diquark-antiquark picture, 
it can be written as~\cite{Wang:2018alb}
\begin{eqnarray}
 J_{\mu}(x)&=&\varepsilon^{abc}\varepsilon^{ade}\varepsilon^{bfg}\big[  q^{d^T}_{1}(x) C\gamma_5 q_2^e(x)\,q_3^{f^T}(x) 
 C\gamma_\mu c^g(x)\, C\bar{s}^{c^T}(x)\big],\nonumber\\
 \label{eq:JJPc}
\end{eqnarray}
%%%%%%%%%%%%%%%%%%%%%%%%%%%%%%%%%%%%
where $q_{1}$, $q_{2}$, $q_{3}$  are u, d and/or s-quark, $C$ is the charge conjugation operator; 
and $a$, $b$... represent color indices.

According to the philosophy of the QCD sum rules, the correlator, given in Eq. (\ref{edmn01}), can be calculated in two ways: 
1) In terms of hadron parameters such
as the masses, residues and the coupling constants, known as hadronic representation;  
2)  in terms of the quark-gluon parameters 
and using the photon DAs which include all nonperturbative
dynamics, known as QCD representation. Then equating these two different 
representations of the correlation function to each other by the help of the quark-hadron duality
assumption gives us the desired sum rules. In order to suppress the contributions of the higher states and 
continuum we apply Borel transformation, 
and continuum subtraction to both sides of the acquired QCD sum rules.

We start to compute the correlation function in terms of hadronic degrees of freedom including the 
physical properties of the particles under consideration. For this purpose,
we insert an intermediate set of $P_{c\bar s}$ pentaquark into the 
correlation function. Consequently, we get
\begin{eqnarray}\label{edmn02}
\Pi^{Had}_{\mu\nu}(p,q)&=&\frac{\langle0\mid J_{\mu}\mid
{P_{c\bar s}}(p)\rangle}{[p^{2}-m_{{P_{c\bar s}}}^{2}]}\langle {P_{c\bar s}}(p)\mid
{P_{c\bar s}}(p+q)\rangle_\gamma\frac{\langle {P_{c\bar s}}(p+q)\mid
\bar{J}_{\nu}\mid 0\rangle}{[(p+q)^{2}-m_{{P_{c\bar s}}}^{2}]},
\end{eqnarray}
%where q is the momentum of the photon. 
The matrix elements in Eq. (\ref{edmn02}) are described as~\cite{Pascalutsa:2006up,Ramalho:2009vc},
\begin{eqnarray}\label{matelpar}
\langle0\mid J_{\mu}\mid {P_{c\bar s}}(p,s)\rangle&=&\lambda_{{P_{c\bar s}}}u_{\mu}(p,s),\nonumber\\
%%%%%%%%%%%%%%%%%%%%%%%%%%%%
\langle {P_{c\bar s}}(p)\mid {P_{c\bar s}}(p+q)\rangle_\gamma &=&-e\bar
u_{\mu}(p)\left\{F_{1}(q^2)g_{\mu\nu}\eslash-
\frac{1}{2m_{{P_{c\bar s}}}}\left
[F_{2}(q^2)g_{\mu\nu}+F_{4}(q^2)\frac{q_{\mu}q_{\nu}}{(2m_{{P_{c\bar s}}})^2}\right]\eslash\qslash
\right.\nonumber\\&+&\left.
F_{3}(q^2)\frac{1}{(2m_{{P_{c\bar s}}})^2}q_{\mu}q_{\nu}\eslash\right\} u_{\nu}(p+q),
\end{eqnarray}
where $\varepsilon$ and q are the polarization vector  and momentum of the photon, respectively,  
$\lambda_{{P_{c\bar s}}}$ denotes the residue 
and $u_{\mu}(p,s)$ is the Rarita-Schwinger spinor of ${P_{c\bar s}}$ pentaquarks. 
Summation on spins of ${P_{c\bar s}}$ pentaquark is performed as:

\begin{align}\label{raritabela}
\sum_{s}u_{\mu}(p,s)\bar u_{\nu}(p,s)=-\Big(\pslash+m_{{P_{c\bar s}}}\Big)\Big[g_{\mu\nu}
-\frac{1}{3}\gamma_{\mu}\gamma_{\nu}-\frac{2\,p_{\mu}p_{\nu}}
{3\,m^{2}_{{P_{c\bar s}}}}+\frac{p_{\mu}\gamma_{\nu}-p_{\nu}\gamma_{\mu}}{3\,m_{{P_{c\bar s}}}}\Big].
\end{align}

In principle, it is possible to acquire the final form of the hadronic representation of the correlator
 using the above equations, but we encounter with two problems: not all Lorentz 
structures are independent and the correlator can include not only the spin-3/2 contributions but also the contributions 
from the spin-1/2 particles, which must be removed. 
To eliminate the spin-1/2 contributions and acquire only independent structures in the
correlator, we order the  Dirac matrices as $\gamma_{\mu}\pslash\eslash\qslash\gamma_{\nu}$ and remove terms starting 
with $\gamma_\mu$, and ending with $\gamma_\nu$ and those which are proportional to $p_\mu$ and 
$p_\nu$~\cite{Belyaev:1982cd}. This procedure eliminates the spin-$\frac{1}{2}$ pollutions.
Consequently, using Eqs. (\ref{edmn02}) and (\ref{matelpar})
for hadronic side we get,
\begin{eqnarray}\label{final phenpart}
\Pi^{Had}_{\mu\nu}(p,q)&=&-\frac{\lambda_{_{{P_{c\bar s}}}}^{2}}{[(p+q)^{2}-m_{_{{P_{c\bar s}}}}^{2}][p^{2}-m_{_{{P_{c\bar s}}}}^{2}]}
\Bigg[  -g_{\mu\nu}\pslash\eslash\qslash \,F_{1}(q^2) 
+m_{{P_{c\bar s}}}g_{\mu\nu}\eslash\qslash\,F_{2}(q^2)+
\frac{F_{3}(q^2)}{4m_{{P_{c\bar s}}}}q_{\mu}q_{\nu}\eslash\qslash\, \nonumber\\&+&
\frac{F_{4}(q^2)}{4m_{{P_{c\bar s}}}^3}(\varepsilon.p)q_{\mu}q_{\nu}\pslash\qslash \,+
\mbox{other independent structures} \Bigg].
\end{eqnarray}

The magnetic dipole $(G_{M}(q^2))$,  electric quadrupole 
$(G_{Q}(q^2))$, and  magnetic octupole $(G_{O}(q^2))$, form factors are
described in terms of the form factors $F_{i}(q^2)$ as
 \cite{Pascalutsa:2006up,Ramalho:2009vc}:
\begin{eqnarray}
G_{M}(q^2) &=& \left[ F_1(q^2) + F_2(q^2)\right] ( 1+ \frac{4}{5}
\lambda ) -\frac{2}{5} \left[ F_3(q^2)  +
F_4(q^2)\right] \lambda \left( 1 + \lambda \right), \nonumber\\
G_{Q}(q^2) &=& \left[ F_1(q^2) -\lambda F_2(q^2) \right]  -
\frac{1}{2}\left[ F_3(q^2) -\lambda F_4(q^2)
\right] \left( 1+ \lambda \right).  \nonumber\\
 G_{O}(q^2) &=&
\left[ F_1(q^2) + F_2(q^2)\right] -\frac{1}{2} \left[ F_3(q^2)  +
F_4(q^2)\right] \left( 1 + \lambda \right),
\end{eqnarray}
  where $\lambda
= -\frac{q^2}{4m^2_{{P_{c\bar s}}}}$. At $q^2=0$, the electromagnetic multipole form factors
are acquired in terms of the functions $F_i(0)$ as:
\begin{eqnarray}\label{mqo1}
G_{M}(0)&=&F_{1}(0)+F_{2}(0),\nonumber\\
G_{Q}(0)&=&F_{1}(0)-\frac{1}{2}F_{3}(0),\nonumber\\
G_{O}(0)&=&F_{1}(0)+F_{2}(0)-\frac{1}{2}[F_{3}(0)+F_{4}(0)].
\end{eqnarray}
The  magnetic dipole, ($\mu_{{P_{c\bar s}}}$),  electric quadrupole
($Q_{{P_{c\bar s}}}$)  and magnetic octupole moments ($O_{{P_{c\bar s}}}$) are described as follows,
 \begin{eqnarray}\label{mqo2}
\mu_{{P_{c\bar s}}}&=&\frac{e}{2m_{{P_{c\bar s}}}}G_{M}(0),\nonumber\\
Q_{{P_{c\bar s}}}&=&\frac{e}{m_{{P_{c\bar s}}}^2}G_{Q}(0),\nonumber\\
O_{{P_{c\bar s}}}&=&\frac{e}{2m_{{P_{c\bar s}}}^3}G_{O}(0).
\end{eqnarray}
In present work we derive sum rules for the form factors $ F_i(q^2) $ then in numerical analyses we will use the above relations to extract the values of the multipole moments using the  sum rules for the form factors. The final form of the hadronic side in terms of the selected structures in momentum space is:
\begin{eqnarray}
\Pi^{Had}_{\mu\nu}(p,q)&=&\Pi_1^{Had}g_{\mu\nu}\pslash\eslash\qslash \,
+\Pi_2^{Had}g_{\mu\nu}\eslash\qslash\,+
\Pi_3^{Had}q_{\mu}q_{\nu}\eslash\qslash\, +
\Pi_4^{Had}(\varepsilon.p)q_{\mu}q_{\nu}\pslash\qslash \,+
...,
\end{eqnarray}
where $ \Pi_i^{Had} $ are functions of the form factors $ F_i(q^2) $ and other hadronic parameters; and $ ... $ represents other independent structures.

In the deep Euclidean region, the correlation function can also be computed in terms of quark-gluon
fields as well as the photon DAs. 
Using expressions of interpolating currents
and contracting all quark pairs, we get the following expression for the correlation function:
\begin{align}
\label{edmn11}
\Pi^{QCD}_{\mu\nu}(p,q)=&i\,\varepsilon^{abc}\varepsilon^{a^{\prime}b^{\prime}c^{\prime}}\varepsilon^{ade}
\varepsilon^{a^{\prime}d^{\prime}e^{\prime}}\varepsilon^{bfg}
\varepsilon^{b^{\prime}f^{\prime}g^{\prime}}\int d^4x e^{ip\cdot x} \langle 0| \widetilde S_s^{c^{\prime}c}(-x)
\Bigg\{ Tr\Big[\gamma_5 \widetilde S_c^{gg^\prime}(x) \gamma_5  S_{q_3}^{ff^\prime}(x)\Big]
Tr\Big[\gamma_\nu \widetilde S_{q_2}^{ee^\prime}(x) \gamma_\mu S_{q_1}^{dd^\prime}(x)\Big] \nonumber\\
&-
Tr\Big[\gamma_5 \widetilde S_c^{gg^\prime}(x) \gamma_5  S_{q_3}^{ff^\prime}(x)\Big]
Tr\Big[\gamma_\nu \widetilde S_{q_1 q_2}^{de^\prime}(x) \gamma_\mu S_{q_2 q_1}^{ed^\prime}(x)\Big]  
%%%%%%%%%%%%%%%%%%%%%%%%%%%%%%%%%%%%%
-  Tr \Big[\gamma_5 \widetilde S_c^{gg^\prime}(x) \gamma_5  S_{q_3 q_1}^{fd^\prime}(x) 
\gamma_\nu \widetilde S_{q_2}^{ee^\prime}(x) \gamma_\mu \widetilde S_{q_1 q_3}^{df^\prime}(x)\Big] \nonumber\\
&-
  Tr \Big[\gamma_5 \widetilde S_c^{gg^\prime}(x) \gamma_5  S_{q_3 q_2}^{fe^\prime}(x) 
\gamma_\nu \widetilde S_{q_1}^{dd^\prime}(x) \gamma_\mu \widetilde S_{q_2 q_3}^{ef^\prime}(x)\Big]
+
 Tr \Big[\gamma_5 \widetilde S_c^{gg^\prime}(x) \gamma_5  S_{q_3 q_1}^{fd^\prime}(x) 
\gamma_\nu \widetilde S_{q_1 q_2}^{de^\prime}(x) \gamma_\mu \widetilde S_{q_3 q_2}^{ef^\prime}(x)\Big]\nonumber\\
&+
Tr \Big[\gamma_5 \widetilde S_c^{gg^\prime}(x) \gamma_5  S_{q_3 q_2}^{fe^\prime}(x) 
\gamma_\nu \widetilde S_{q_2 q_1}^{ed^\prime}(x) \gamma_\mu \widetilde S_{q_1 q_3}^{df^\prime}(x)\Big]
\Bigg \}
|0 \rangle_\gamma,
\end{align}
 where 
$ \widetilde{S}_{c(q)}^{ij}(x)=CS_{c(q)}^{ij\mathrm{T}}(x)C$ and $S_{q_{i}q_{j}}$ exists when $q_i=q_j$
but it vanishes when $q_i\neq q_j$.

%with $S_{q(c)}(x)$ being the quark propagator.
The quark propagators $S_{q}(x)$ and $S_{c}(x)$ are given as~\cite{Balitsky:1987bk}
\begin{eqnarray}
\label{edmn12}
S_{q}(x)=S^{free}_q
- \frac{ \bar qq }{12} \Big(1-i\frac{m_{q} \xslash}{4}   \Big)
- \frac{ \bar q \sigma.G q }{192} x^2  \Big(1-i\frac{m_{q} \xslash}{6}   \Big)
-\frac {i g_s }{32 \pi^2 x^2} ~G^{\mu \nu} (x) \Bigg[\rlap/{x} 
\sigma_{\mu \nu} +  \sigma_{\mu \nu} \rlap/{x}
 \Bigg],
\end{eqnarray}%
and
\begin{eqnarray}
\label{edmn13}
S_{c}(x)=S^{free}_c
-\frac{g_{s}m_{c}}{16\pi ^{2}} \int_{0}^{1}dv~G^{\mu \nu }(vx)\Bigg[ (\sigma _{\mu \nu }{\xslash}
  +{\xslash}\sigma _{\mu \nu })\frac{K_{1}( m_{c}\sqrt{-x^{2}}) }{\sqrt{-x^{2}}}
+2\sigma ^{\mu \nu }K_{0}( m_{c}\sqrt{-x^{2}})\Bigg],
\end{eqnarray}%
where 
\begin{eqnarray}
&&S_{q}^{free}(x)=i \frac{{\xslash}}{2\pi ^{2}x^{4}} -\frac{m_{q}}{4 \pi^2 x^2},\nonumber\\
\nonumber\\
 &&S_{c}^{free}(x)=\frac{m_{c}^{2}}{4 \pi^{2}} \Bigg[ \frac{K_{1}(m_{c}\sqrt{-x^{2}}) }{\sqrt{-x^{2}}}
+i\frac{{\xslash}~K_{2}( m_{c}\sqrt{-x^{2}})}
{(\sqrt{-x^{2}})^{2}}\Bigg],
\end{eqnarray}
with $K_i$ being the  modified  Bessel functions of the second kind.

The correlation function includes short distance (perturbative), and long distance (nonperturbative)
contributions.
In the first part,  the propagator of the quark interacting with the photon perturbatively is replaced by
\begin{align}
\label{free}
S^{free}(x) \rightarrow \int d^4y\, S^{free} (x-y)\,\rlap/{\!A}(y)\, S^{free} (y)\,,
\end{align}
and the remaining four propagators in Eq.~(\ref{edmn11}) are replaced with the full quark propagators
including the perturbative and  nonperturbative parts.  Here we use $ A_\mu(y)=-\frac{1}{2}\, F_{\mu\nu}(y)\, y^\nu $ where  the electromagnetic field strength tensor is written as $ F_{\mu\nu}(y)=-i(\varepsilon_\mu q_\nu-\varepsilon_\nu q_\mu)\,e^{iq.y} $.
The total perturbative contribution is acquired by performing the   
replacement mentioned above for the perturbatively interacting quark propagator with the photon and   
making use of the replacement of the remaining propagators by their free parts.

In the next part, one of the light quark propagators in Eq.~(\ref{edmn11}), 
defining the photon emission at large distances, is substitute by
\begin{align}
\label{edmn14}
S_{\alpha\beta}^{ab}(x) \rightarrow -\frac{1}{4} \big[\bar{q}^a(x) \Gamma_i q^b(x)\big]\big(\Gamma_i\big)_{\alpha\beta},
\end{align}
and the rest propagators are substituted with the full quark propagators.
 Here, $\Gamma_i$ represent the full set of Dirac matrices. Once 
Eq. (\ref{edmn14}) is plugged into Eq. (\ref{edmn11}), there appear matrix
elements of  $\langle \gamma(q)\vel \bar{q}(x) \Gamma_i q(0) \ver 0\rangle$
and $\langle \gamma(q)\vel \bar{q}(x) \Gamma_i G_{\alpha\beta}q(0) \ver 0\rangle$ kinds,
representing the nonperturbative contributions. 
To calculate the nonperturbative contributions, we need these matrix elements which are parameterized 
in terms of photon wave functions with definite twists. The explicit expressions of the photon DAs are presented in Ref.~\cite{Ball:2002ps}.
The QCD side of the correlation function can be acquired in terms of quark-gluon parameters 
as well as the DAs of the photon 
 using Eqs.~(\ref{edmn11})-(\ref{edmn14}) and after performing the Fourier transformation to 
remove the calculations to the momentum space.
As a result of above procedures the QCD side of the correlation function in terms of the selected structures in momentum space is obtained as
\begin{eqnarray}
\Pi^{QCD}_{\mu\nu}(p,q)&=&\Pi_1^{QCD}g_{\mu\nu}\pslash\eslash\qslash \,
+\Pi_2^{QCD}g_{\mu\nu}\eslash\qslash\,+
\Pi_3^{QCD}q_{\mu}q_{\nu}\eslash\qslash\, +
\Pi_4^{QCD}(\varepsilon.p)q_{\mu}q_{\nu}\pslash\qslash \,+
...,
\end{eqnarray}
where $ \Pi_i^{QCD} $ are functions of the QCD degrees of freedom and photon DAs parameters.

The sum rules are obtained by equating the hadronic
and QCD representations of the correlation function.
The next step  is to  perform
double Borel transformation (${\cal B}  $)
over the $p^2$ and $(p+q)^2$ on the both sides of the sum rules 
in order to stamp down the contributions of higher states and continuum. To further suppress the contributions of the higher states and continuum we apply the continuum subtraction and use the quark-hadron duality assumption.  Hence,
\begin{eqnarray}
{\cal B}\Pi^{Had}_{\mu\nu}(p,q)={\cal B}\Pi^{QCD}_{\mu\nu}(p,q),
\end{eqnarray}
which leads to
\begin{eqnarray}
{\cal B}\Pi_1^{Had}={\cal B}\Pi_1^{QCD},~~~~{\cal B}\Pi_2^{Had}={\cal B}\Pi_2^{QCD},~~~~{\cal B}\Pi_3^{Had}={\cal B}\Pi_3^{QCD},~~~~{\cal B}\Pi_4^{Had}={\cal B}\Pi_4^{QCD},
\end{eqnarray}
 corresponding to the structures 
$g_{\mu\nu}\pslash\eslash\qslash$, $g_{\mu\nu}\eslash\qslash$, $q_{\mu}q_{\nu}\eslash\qslash$ and  
$(\varepsilon.p)q_{\mu}q_{\nu}\pslash\qslash$. By this way we obtain the sum rules for the form factors 
$F_1$, $F_2$, $F_3$ and $F_4$, whose explicit expressions  are presented in the Appendix.

\section{Numerical analysis}

This section is devoted to the numerical analysis of the electromagnetic multipole moments
of the charm-strange $P_{c\bar s}$ pentaquarks. 
We use $m_u=m_d=0$, $m_s ~(2~ GeV) = 0.096^{+0.08}_{-0.04}~GeV$, $\overline{m_c} (m_c) = (1.28\pm 0.03)\,GeV$~(in $\overline{MS}$ scheme)~\cite{Patrignani:2016xqp}, 
$f_{3\gamma}(\mu = 1~ GeV) =-0.0039~GeV^2$~\cite{Ball:2002ps}, 
%$m_b = (4.78\pm 0.06)\,GeV$, 
$\langle \bar uu\rangle (\mu = 1~GeV) = 
\langle \bar dd\rangle (\mu = 1~GeV)=(-0.24\pm0.01)^3\,GeV^3$ \cite{Ioffe:2005ym}, 
$ \langle \bar ss \rangle (\mu = 1~ GeV) = 0.8 \langle \bar uu \rangle (\mu = 1~ GeV) $,
$m_0^{2}~ (\mu = 1~ GeV) = 0.8 \pm 0.1~GeV^2$, 
%and 
$\langle g_s^2G^2\rangle = 0.88~ GeV^4$~\cite{Nielsen:2009uh}
and $\chi(\mu = 1~ GeV)=-2.85 \pm 0.5~GeV^{-2}$~\cite{Rohrwild:2007yt}. 
To obtain a numerical values for the electromagnetic form factors, 
we need to determine the values of the mass and residue of the $P_{c\bar s}$ pentaquarks.
The mass and residue of the $P_{c\bar s}$ pentaquarks are borrowed from \cite{Wang:2018alb}.
The parameters entering the photon DAs are presented in Ref.~\cite{Ball:2002ps}.

The estimations for the electromagnetic multipole moments of the charm-strange $P_{c\bar s}$ pentaquarks 
depend on two auxiliary parameters; the continuum
threshold $s_0$ and Borel mass parameter $M^2$. 
In order to obtain reliable values
of the electromagnetic multipole moments from QCD sum rules, 
we should find the working regions of $s_0$ and $M^2$  in
such a way that the results are insensitive to the variation of these parameters.  
To obtain a working region for $M^2$, we require the pole dominance over the contributions of higher
states and continuum. And also the results coming from higher dimensional operators should contribute less than the
lower dimensional ones, since operator product expansion (OPE) should be convergent.
The above requirements restrict the working
region of the Borel parameter to $3~GeV^2 \leq M^2 \leq 5~GeV^2$. 
The continuum threshold $s_0$
is not totally arbitrary and it is relevant to the
energy of the first corresponding excited state. In its fixing
we again consider the OPE convergence and pole dominance.
 Our numerical calculations lead to the interval [11-13] $GeV^2$
for this parameter. 
%%%
 In Fig. 1, as example, we plot the dependencies of the magnetic dipole moments of the possible pentaquarks on $M^2$ 
 at several fixed values of the continuum threshold $s_0$.  
 From these graphics we observe that the corresponding magnetic dipole moments
 seem to be almost independent of  $M^2$  for different choices of $s_0$. 
However, the dependencies of the obtained results on the continuum 
threshold are considerable eventhough they are within the limits allowed by the standard prescriptions of the method. 
We include these variations in the errors of our final results.

 \begin{figure}
\centering
 \includegraphics[width=0.31\textwidth]{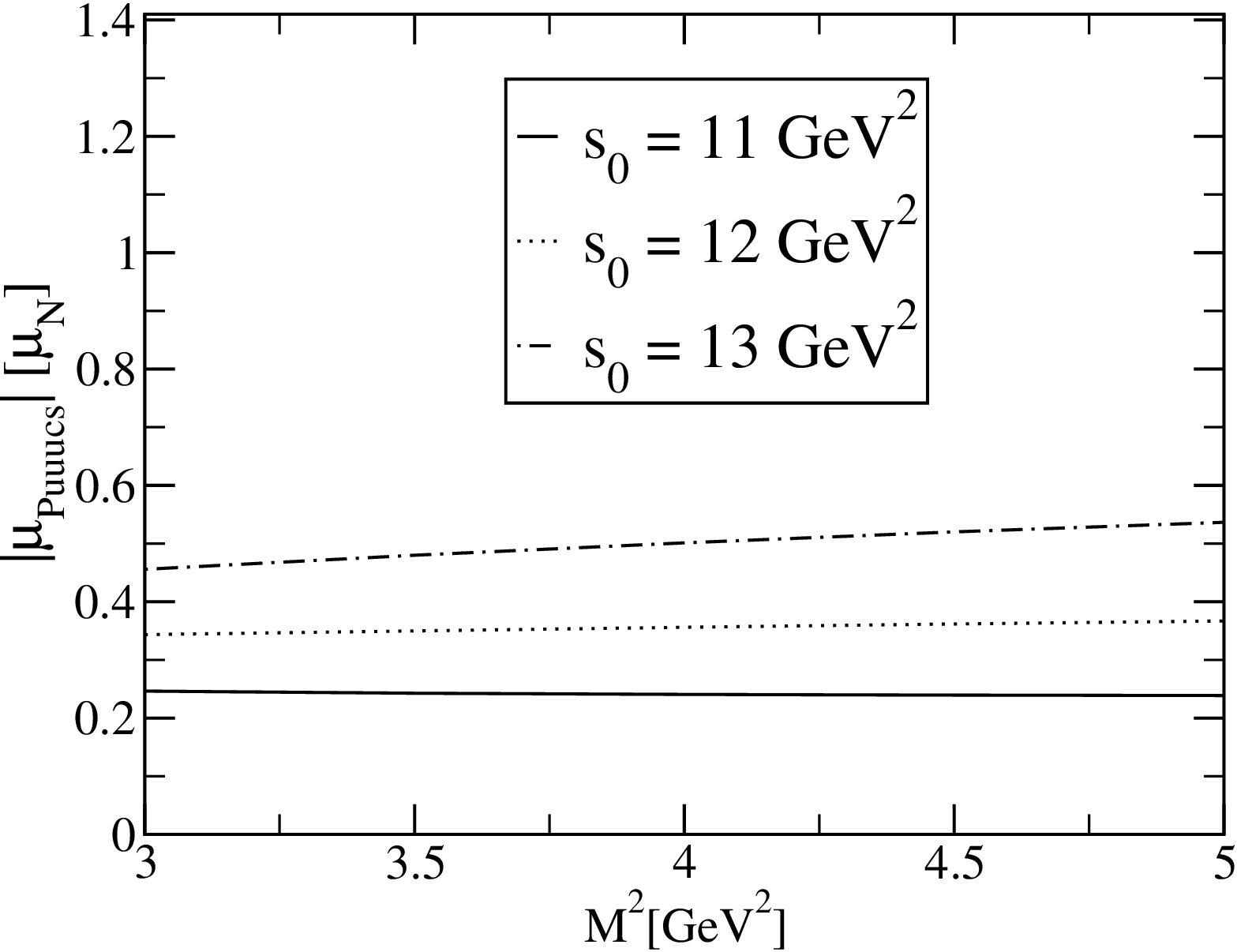}~~~
 \includegraphics[width=0.31\textwidth]{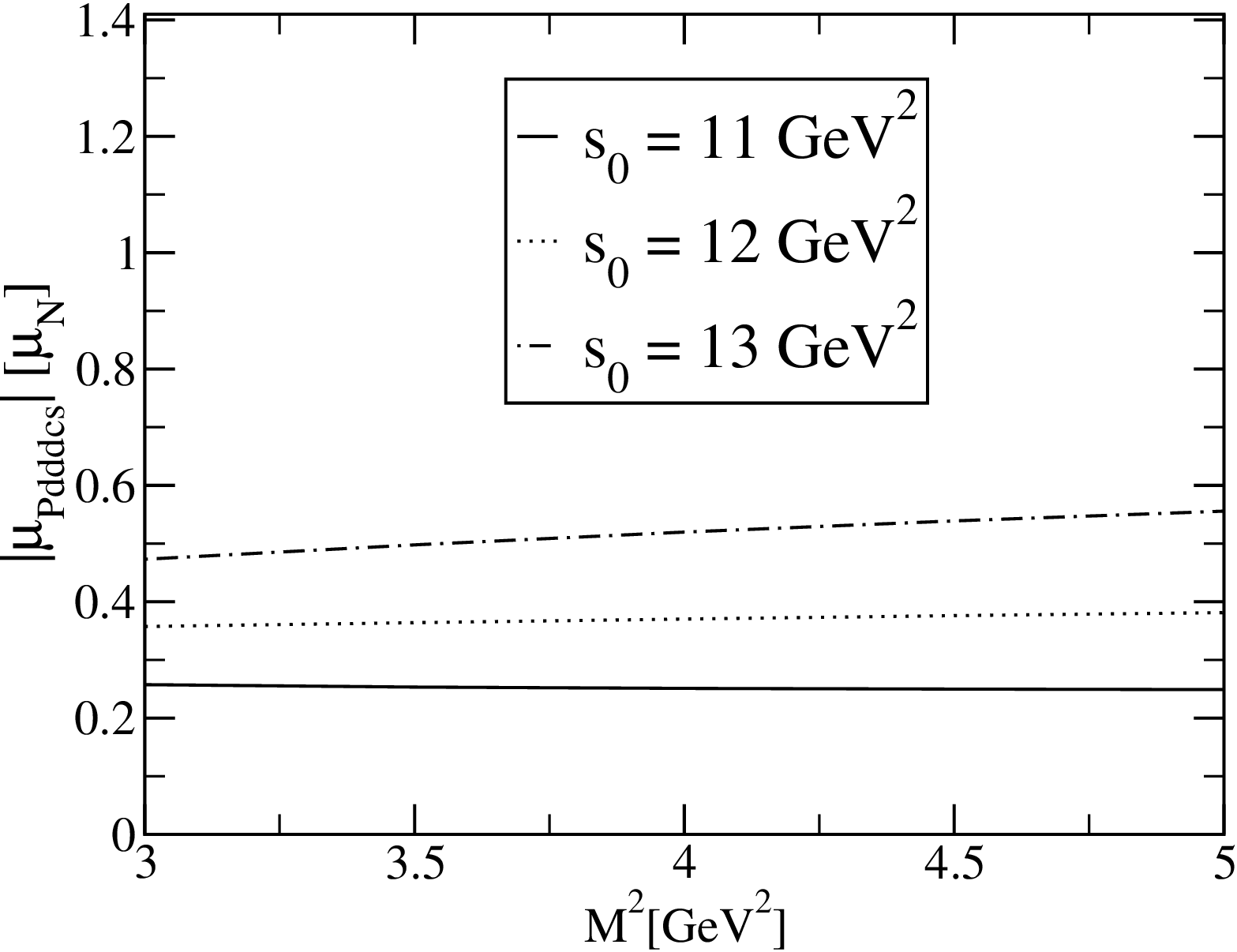}~~~
  \includegraphics[width=0.31\textwidth]{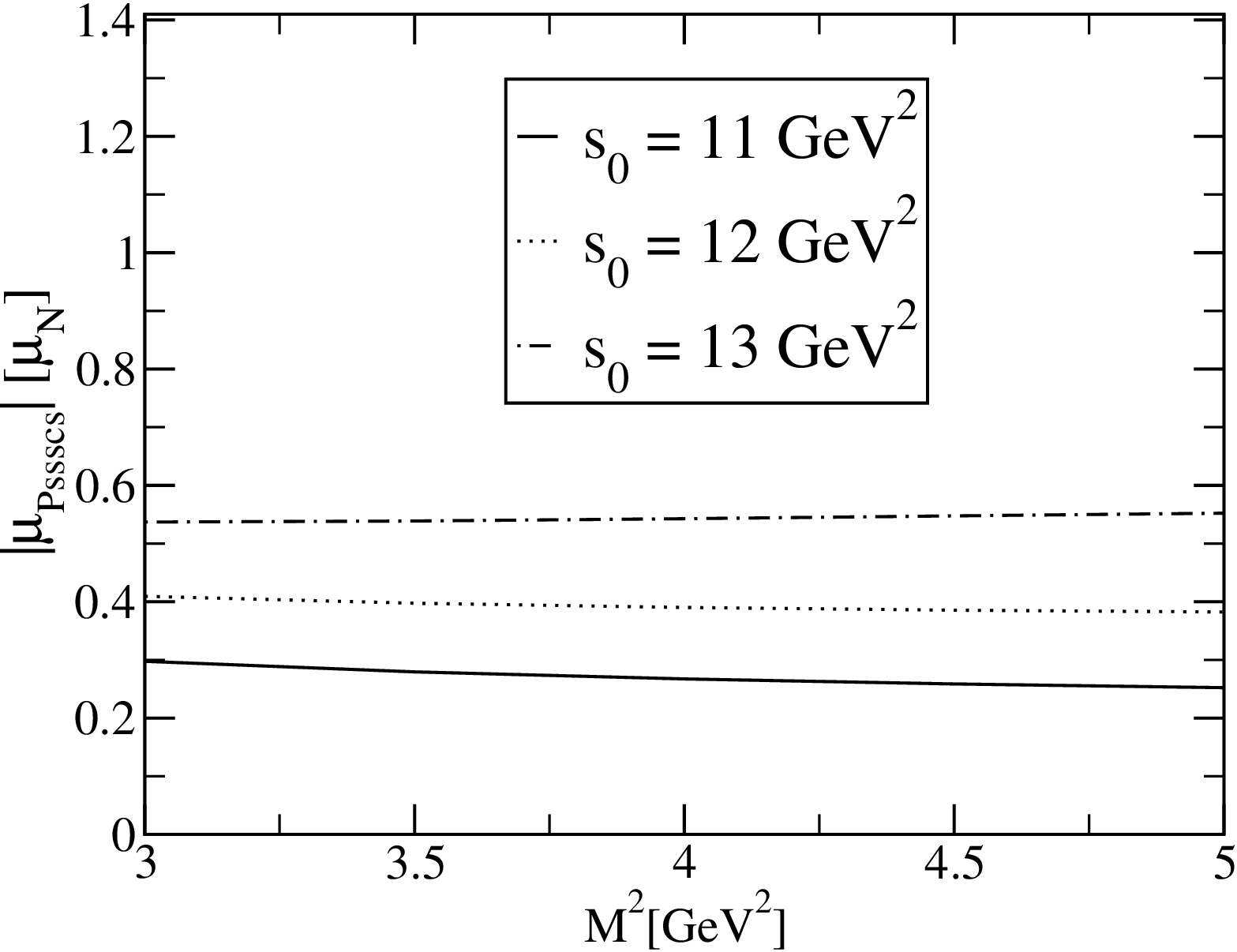}\\
 \vspace{0.8cm}
  \includegraphics[width=0.31\textwidth]{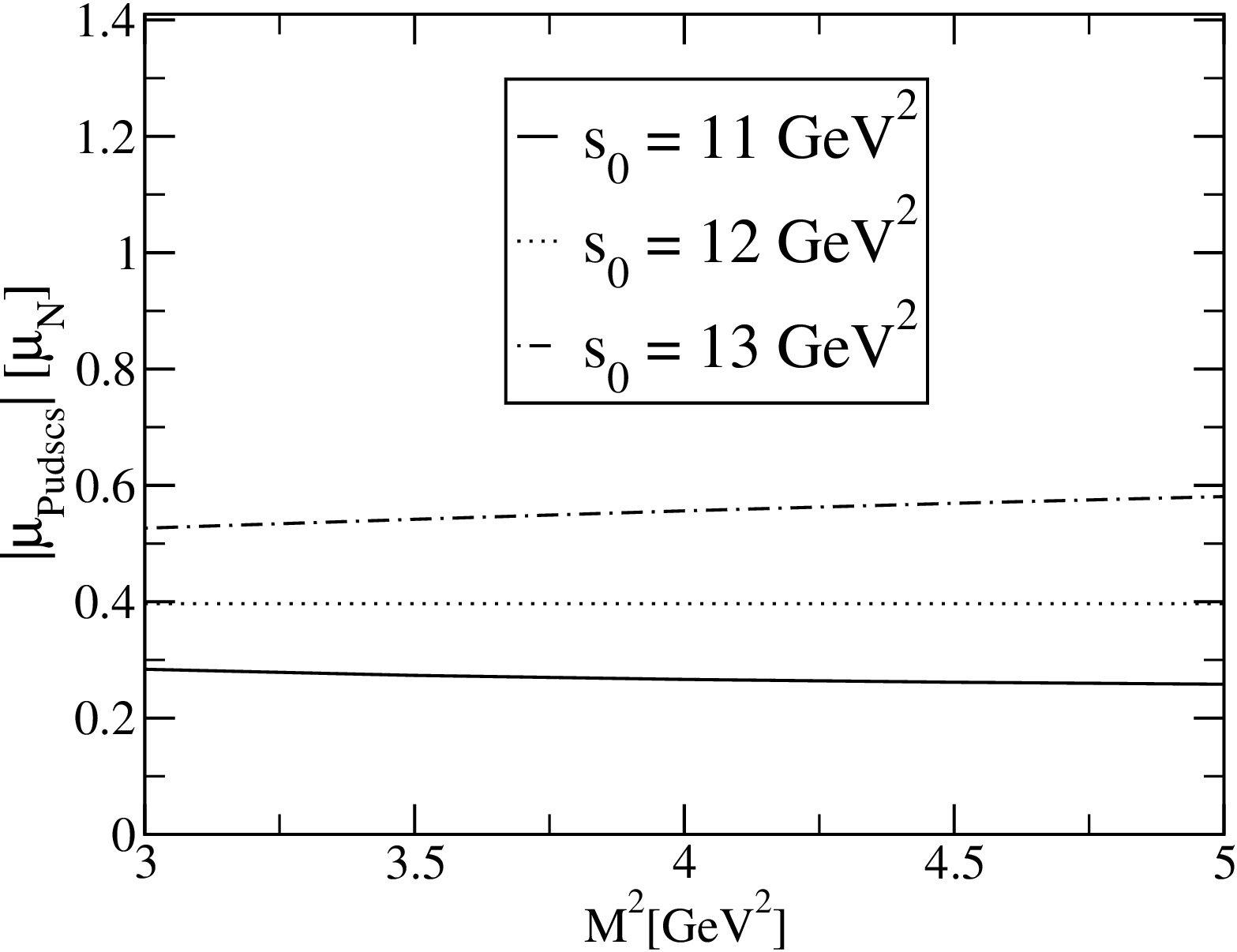}~~~
  \includegraphics[width=0.31\textwidth]{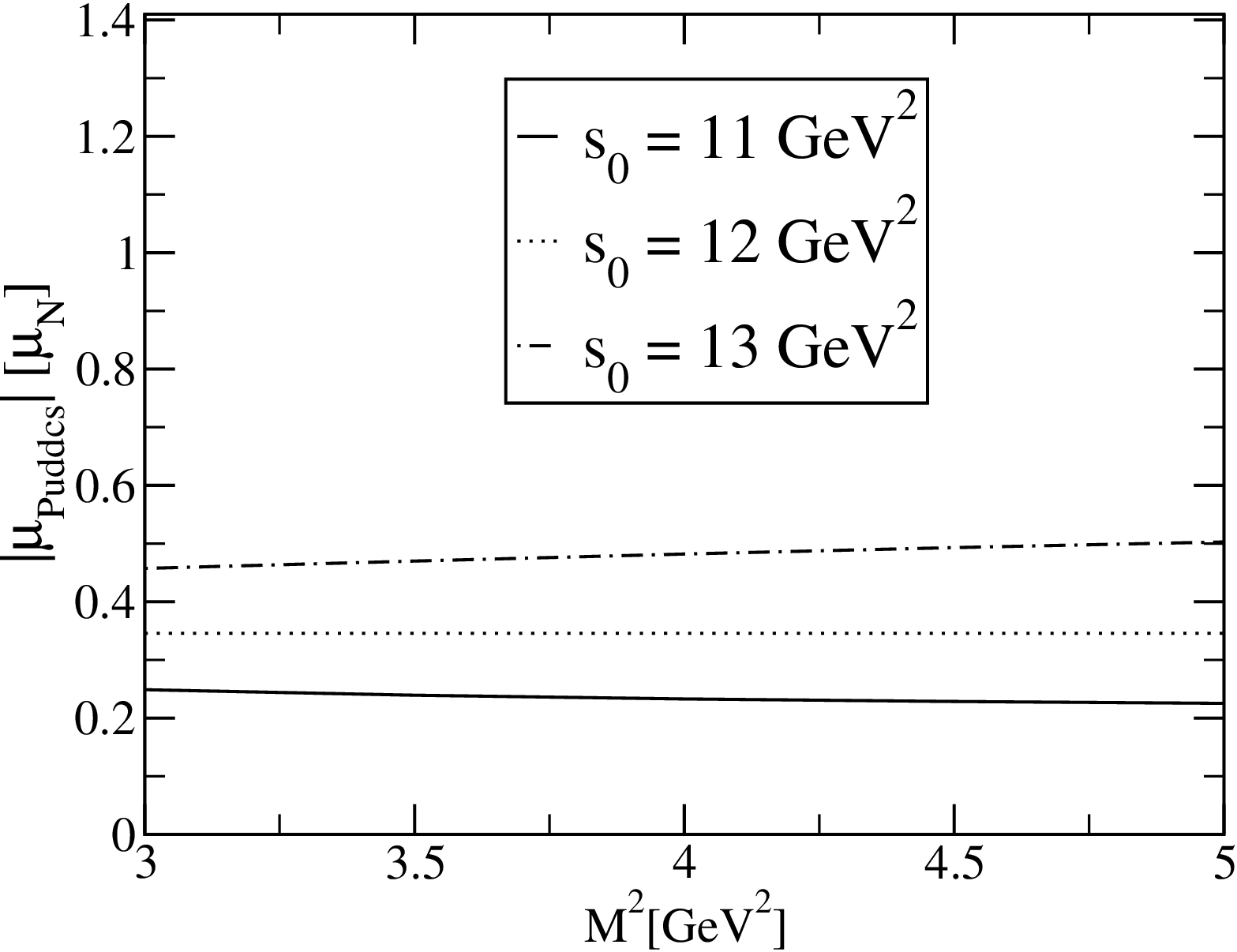}~~~
  \includegraphics[width=0.31\textwidth]{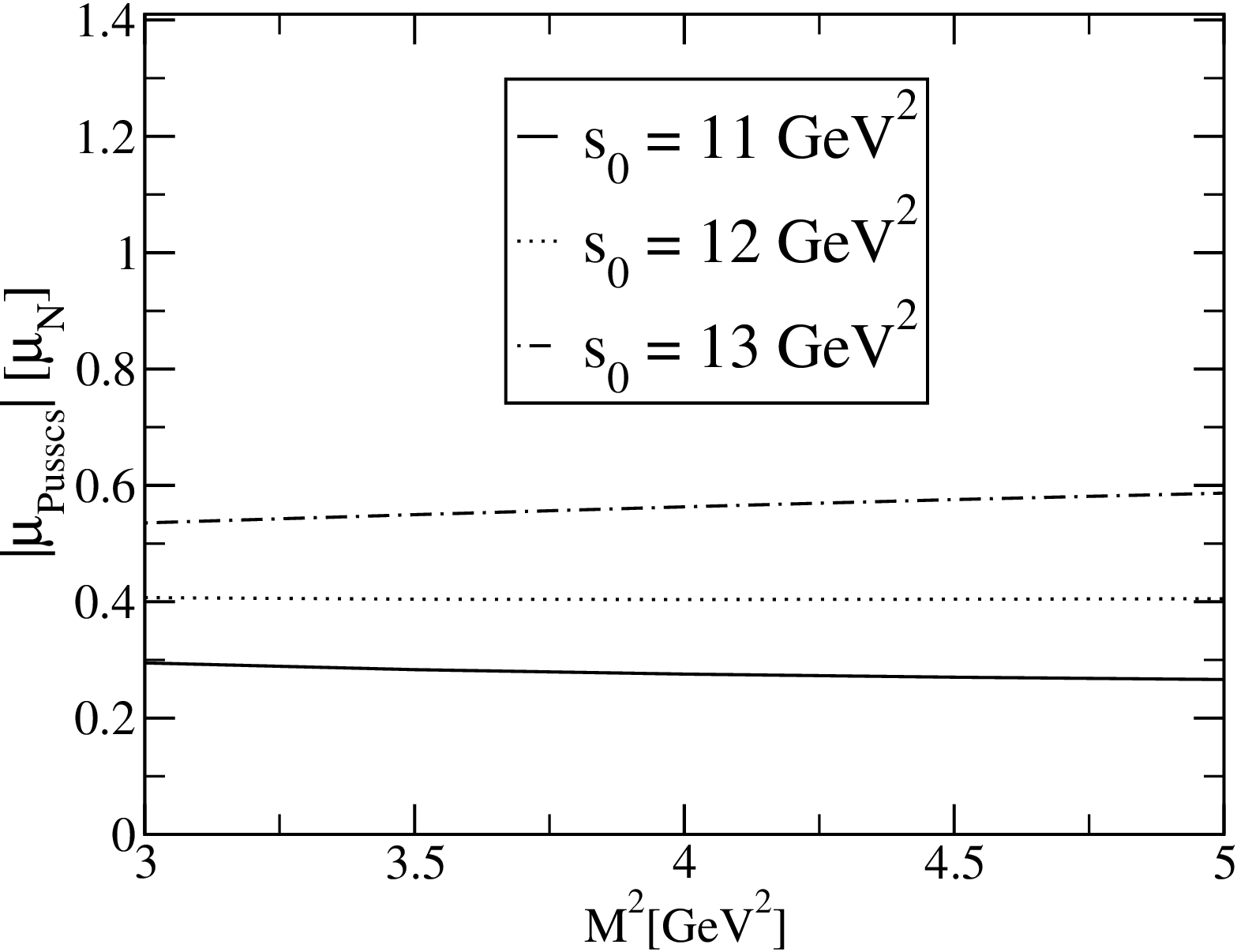}\\
  \vspace{0.8cm}
  \includegraphics[width=0.31\textwidth]{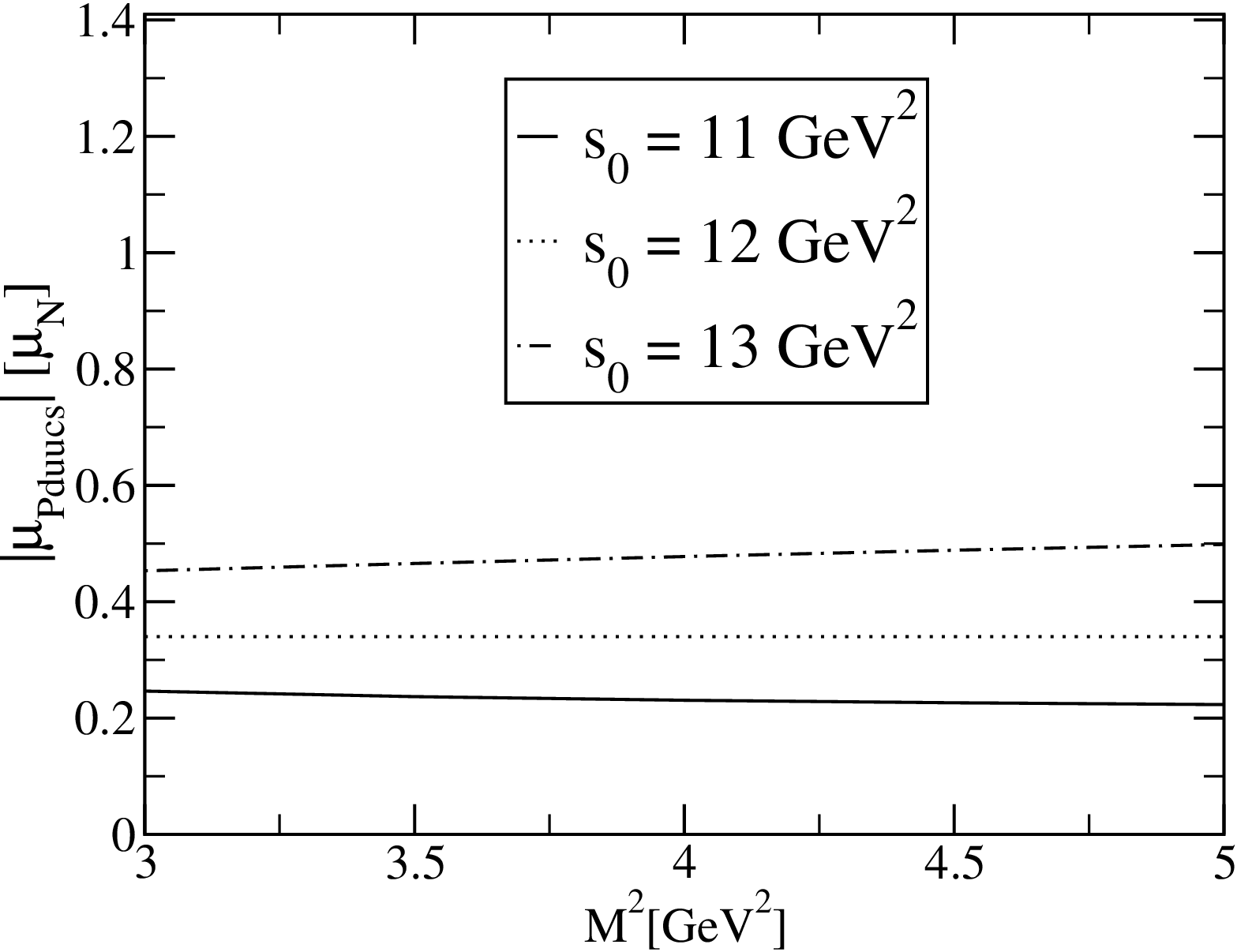}~~~
  \includegraphics[width=0.31\textwidth]{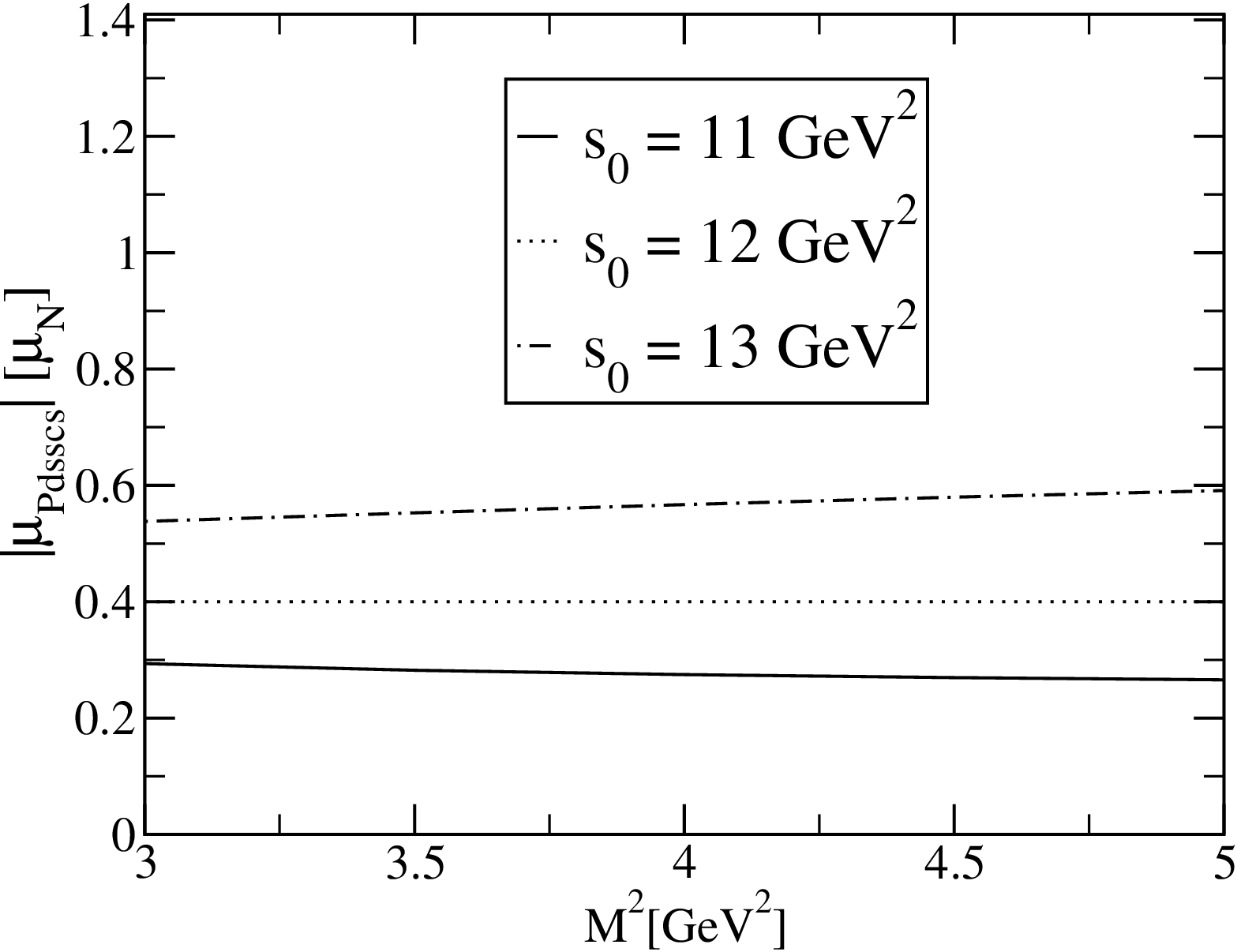}~~~
  \includegraphics[width=0.31\textwidth]{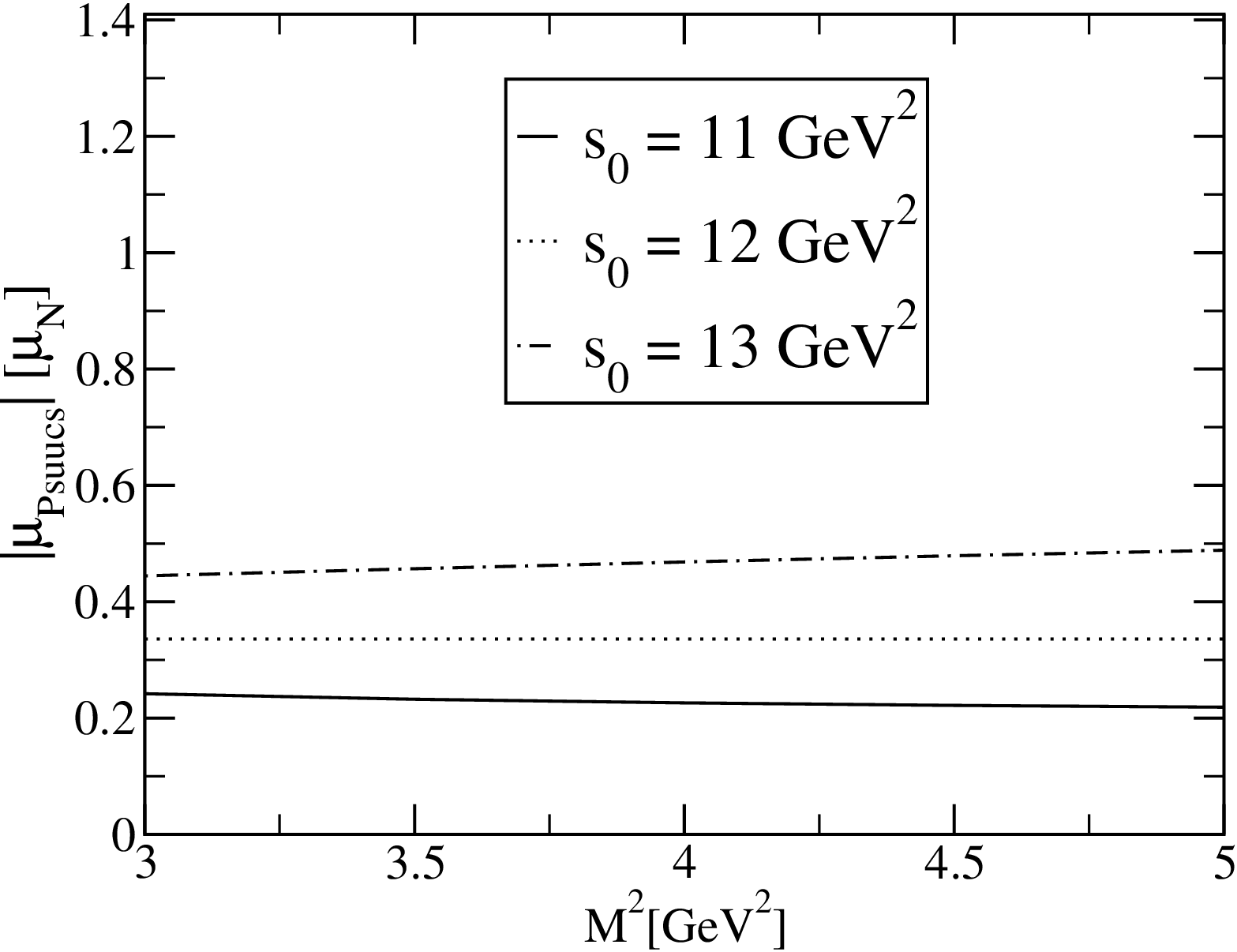}\\
  \vspace{0.8cm}
  \includegraphics[width=0.31\textwidth]{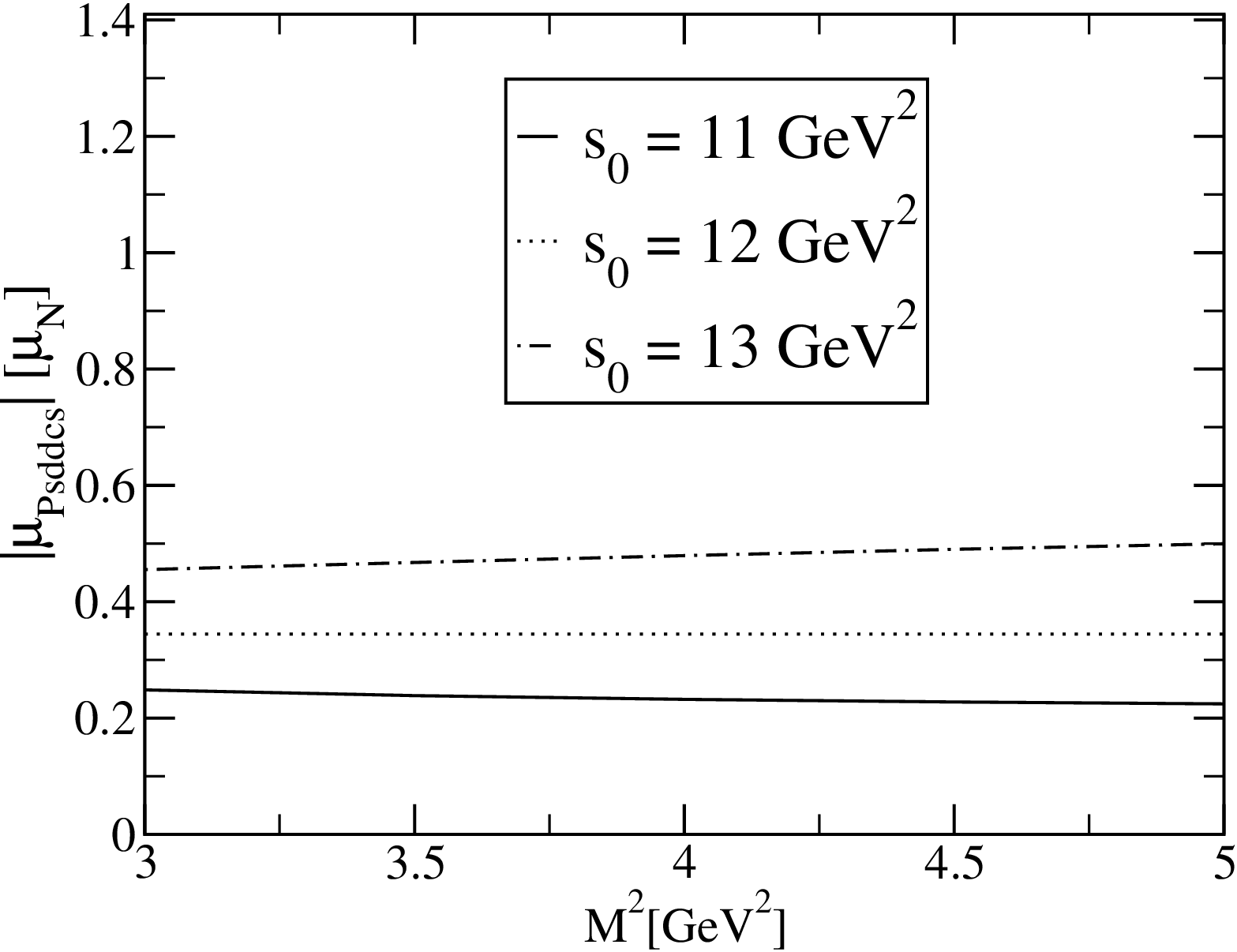}
 \caption{The magnetic dipole moments for $P_{c\bar s}$ pentaquarks versus $M^{2}$ at various fixed values of the $s_0$.}
  \end{figure}

Our results for the magnetic dipole and electric quadrupole moments are shown in Table I.
The magnetic octupole moments of the charm-strange $P_{c\bar s}$ pentaquarks have also been 
calculated but they are not presented here because their values are very close to zero.
The errors in the results are due to the uncertainties carried by the input parameters
and photon DAs as well as those coming from the working windows for auxiliary parameters. 
We should note that the primary source of uncertainties is because of the variations of the results  with respect to $s_0$. 
It is worth mentioning that in Table I and Fig. 1, the absolute values of the quantities 
are shown since it is not possible to define the sign of the residue from
the mass sum rules.  Hence, we cannot predict
the signs of the electromagnetic multipole moments.

\begin{table}[htp]
	\addtolength{\tabcolsep}{10pt}
	\begin{center}
\begin{tabular}{c|c|ccc}
	   \hline\hline
	   State&  $|\mu_{P_{c\bar s}}|[\mu_N]$	& $|Q_{P_{c\bar s}}|[fm^2](\times 10^{-3})$\\
	   \hline\hline
	   $uddc\bar s$&  $0.36 \pm 0.14$      &$0.21 \pm 0.03$ \\
	   $duuc\bar s$&  $0.36 \pm 0.14$      &$0.21 \pm 0.03$ \\
	   $suuc\bar s$&  $0.36 \pm 0.14$      &$0.21 \pm 0.03$ \\
	   $sddc\bar s$&  $0.36 \pm 0.14$      &$0.21 \pm 0.03$ \\
	   $uuuc\bar s$&  $0.39 \pm 0.15$      &$0.22 \pm 0.03$ \\	
	   $dddc\bar s$&  $0.40 \pm 0.15$      &$0.23 \pm 0.04$ \\	
	   $sssc\bar s$&  $0.40 \pm 0.15$      &$0.23 \pm 0.04$ \\
	   $udsc\bar s$&  $0.42 \pm 0.16$      &$0.23 \pm 0.04$ \\
	   $ussc\bar s$&  $0.43 \pm 0.16$      &$0.23 \pm 0.04$ \\
	   $dssc\bar s$&  $0.43 \pm 0.16$      &$0.23 \pm 0.04$ \\
	   	   \hline\hline
\end{tabular}
\end{center}
\caption{Numerical values of the magnetic dipole and  electric quadrupole %and magnetic octupole
moments of $P_{c\bar s}$ pentaquarks.}
	\label{table}
\end{table}
\section{Discussion and Concluding Remarks}
The electromagnetic multipole moments of the charm-strange $P_{c\bar s}$ pentaquarks have been investigated by assuming
that these states are represented in diquark-diquark-antiquark picture with quantum numbers $J^{P } =\frac{3}{2}^-$.
Their magnetic dipole   and electric quadrupole moments have been extracted in the framework of light-cone QCD sum rule. 
The electromagnetic multipole moments of the charm-strange $P_{c\bar s}$ pentaquarks are essential dynamical observables, 
which can contain valuable information of their substructure, charge distribution inside them and their geometric shapes. 
The numerical values obtained for the magnetic dipole moments are large enough to be measured in future experiments.
However we got very small results for the electric quadrupole moments of charm-strange $P_{c\bar s}$ pentaquarks indicating a 
nonspherical charge distribution. As we mentioned above, the values of magnetic octupole 
moments are obtained to be very close to zero.
   
\section{Acknowledgements}
The support of TUBITAK through the Grant No. 115F183 is appreciated.

\section*{QCD sum rules for the electromagnetic form factors \texorpdfstring{$F_i$}{}}
The explicit expressions for the electromagnetic form factors $F_i$ are given as: 
\begin{align}
 F_1&=-\frac{e^{m^2_{P_{c \bar s}}/M^2}}{\lambda_{P_{c \bar s}}^2}\Bigg\{
 \frac {e_s} {1228800\, m_c^2\,\pi^8}\Bigg[
   m_c^5 \Bigg (-\, m_c^{11} \, I[-8, 5] + 5\, m_ {q_ 3} m_c^{10} \, 
      I[-7, 4] + 10\, 
      m_c^7 (-5\, m_ {q_ 1} m_ {q_ 2}\, I[-6, 4] + I[-6, 5])\nonumber\\
      &+ 
       50\, m_ {q_ 3}\, 
       m_c^6 \, \Big (4\, m_ {q_ 1}\, m_ {q_ 2}\, 
         I[-5, 3] - I[-5, 4]\Big) + 10\, 
       m_c^5\, \Big (15\, m_ {q_ 1}\, m_ {q_ 2}\, I[-5, 4] + 2\, 
         I[-5, 5]\Big) + 100\, 
      m_ 3\, m_c^4\, \Big (6\, m_ {q_ 1} m_ {q_ 2}\nonumber\\ 
     &\times    I[-4, 3] + I[-4, 4]\Big) + 15\, 
       m_c^3\, \Big (-10 \, m_ {q_ 1}\, m_ {q_ 2}\, 
         I[-4, 4] + I[-4, 5]) + 75\, 
      m_ {q_ 3}\, m_c^2\, \Big (8 m_{q_1} m_{q_2}\, I[-3, 3] - I[-3, 4]\Big)\nonumber\\
      &+
        m_c \, \Big (50\, m_ {q_ 1}\, m_ {q_ 2}\, I[-3, 4] + 4\, 
         I[-3, 5]\Big) + 20\, 
      m_ {q_ 3}\, \Big (10 \, m_ {q_ 1}\, m_ {q_ 2}\, 
        I[-2, 3] + I[-2, 4]\Big)\Bigg) + 1600\, m_ {q_ 1} \, m_ {q_ 2}\, 
   m_ {q_ 3}\, m_c \, I[0, 3] \nonumber\\
   &+ 48\, I[0, 5]\Bigg]\nonumber\\
   %%%%%%%%%%%%%%%%%%%%%%%%%%%%%%%%%%%%%%%%%%%%%%%%%%%%%%%%
 &-\frac{e_c\,m_c\, \langle g_s^2G^2\rangle }{4718592\, \pi^8}
  \Bigg[6\,m_c\,m_ {q_ {23}}  (m_ {q_ 1} - 
       m_ {q_ {12}})  \big(m_c^4 I[-4, 2] - 
       2 m_c^2 I[-3, 2] + I[-2, 2]\big) \nonumber\\
       &+ 
    m_ {q_ {13}}^2 \Big(m_c^7 \,I[-5, 2] - 3 m_c^3 \,I[-3, 2] + 
       2 m_c\, I[-2, 2]\Big) + 
    m_ {q_ {23}} ^2 \big(m_c^7\, I[-5, 2] - 3\, m_c^3\, I[-3, 2] + 2 m_c\, I[-2, 2]\big) \nonumber\\
    &- 
    2\, m_ {q_ {13}}\, m_c \Bigg(3\, (m_ {q_ {12}} - m_ {q_ {2}}) \Big(m_c^4\, I[-4, 2] - 2\, m_c^2\, I[-3, 2] + 
          I[-2, 2]\Big) + 
       m_ {q_ {23}} \Big(m_c^6\, I[-5, 2] - 3\, m_c^2\, I[-3, 2] \nonumber\\
       &+ 2 \, I[-2, 2]\Big)\Bigg) - 
    2\, (m_ {q_ {1}} - 2\, m_ {q_ {12}} +m_ {q_ {2}}) \Big(m_c^6\, I[-4, 2] - 3\, m_c^4\, I[-3, 2] 
    + 3\, m_c^2\, I[-2, 2]- I[-1, 2]\Big)\Bigg]\nonumber  \\
    %%%%%%%%%%%%%%%%%%%%%%%%%%%
  &  +\frac{e_{q_1}\, \langle g_s^2G^2\rangle}{84934656\, m_c^2\, \pi^8}
  \Bigg[m_c^3\, \Bigg (m_c\, \bigg (m_c^2\, \Big(\,31\, m_c^6 \,I[-6, 3] - 
            6\, m_c^4\, \big(6\, m_{q_ {23}}^2 + 9\, m_ {q_ {23}}\,m_c + 8\, m_ {q_ {3}}\, m_c\big)\, I[-5, 2]
            -             102\, m_c^2\,\nonumber\\
            & \times I[-4, 3]+ 
            36\, (3\, m_{q_ {23}}^2 + 6\, m_{q_ {23}}\, m_c + 4\, m_{q_ {3}}\, m_c)\, I[-3, 2] - 
            80\, I[-3, 3]\Big) - 
         24\, (3\, m_{q_ {23}}^2 + 9\, m_{q_ {23}}\, m_c + 4\, m_{q_ {3}}\, m_c)  \nonumber\\
         &\times I[-2, 2]- 
         9 I[-2, 3]\bigg) + 54\, m_{q_ {23}}\, I[-1, 2]\Bigg) - 160\, I[0, 3]\Bigg]\nonumber\\
         %%%%%%%%%%%%%%%%%%%%%%%%%%%%%%%%%%%%%%%%%%%
         &  +\frac{e_{q_2}\, \langle g_s^2G^2\rangle}{84934656\, m_c^2\, \pi^8}
  \Bigg[m_c^3\, \Bigg (m_c\, \bigg (m_c^2\, \Big(\,31\, m_c^6 \,I[-6, 3] - 
            6\, m_c^4\, \big(6\, m_{q_ {13}}^2 + 9\, m_ {q_ {13}}\,m_c + 8\, m_ {q_ {3}}\, m_c\big)\, I[-5, 2]
            -             102\, m_c^2\,\nonumber\\
            & \times I[-4, 3]+ 
            36\, (3\, m_{q_ {13}}^2 + 6\, m_{q_ {13}}\, m_c + 4\, m_{q_ {3}}\, m_c)\, I[-3, 2] - 
            80\, I[-3, 3]\Big) - 
         24\, (3\, m_{q_ {13}}^2 + 9\, m_{q_ {13}}\, m_c + 4\, m_{q_ {3}}\, m_c)  \nonumber\\
         &\times I[-2, 2]- 
         9 I[-2, 3]\bigg) + 54\, m_{q_ {13}}\, I[-1, 2]\Bigg) - 160\, I[0, 3]\Bigg]\nonumber\\
         %%%%%%%%%%%%%%%%%%%%%%%%%%%%%%%%%%%%%%
       &+\frac{e_{q_3}\, \langle g_s^2G^2\rangle}{5308416\, m_c^2\, \pi^8}
  \Bigg[ m_c^{12}\, I[-6, 3] - 
 6\, m_c^8\, \Big(\,3\, \big(m_{q_ {12}}^2 + m_{q_ {1}}\, m_{q_ {2}}\big) I[-4, 2] + I[-4, 3]\Big) + 
 4\, m_c^6\, \Big(\,9\, \big(m_{q_ {12}}^2 + m_{q_ {1}}\, m_{q_ {2}}\big)\nonumber\\
 &\times I[-3, 2] - 2\, I[-3, 3]\Big) - 
 3\, m_c^4 \,\Big(\,6\, \big(m_{q_ {12}}^2 + m_{q_ {1}}\, m_{q_ {2}}\big) I[-2, 2] + I[-2, 3]\Big) - 
 16 \,I[0, 3]\Bigg]\nonumber\\
 %%%%%%%%%%%%%%%%%%%%%%%%%%%%%%%%%%%%%%%%%%%%%%%%%%%%%%
  &+\frac{e_{q_{12}}\, \langle g_s^2G^2\rangle}{42467328\, m_c^2\, \pi^8}
  \Bigg[ m_c^3 \Bigg (m_c \bigg (m_c^2 \Big (m_c^6 \,I[-6, 3] + 
          3\, m_c^4 \big (12\, m_{q_ {13}}\, m_{q_ {13}} + 9 \,m_{q_ {13}}\, m_c + 9\, m_{q_ {23}}\, m_c - 
             16\, m_{q_ {3}}\, m_c\big)\nonumber\\
             &\times I[-5, 2] + 6\, m_c^2\, I[-4, 3] - 
          36\, \big (\,(3\, m_{q_ {23}} - 4\,m_{q_ {3}}\,) \,m_c + 
             3\, m_{q_ {13}}\, (m_{q_ {23}} + m_c)\big) I[-3, 2] + 
          16\, I[-3, 3]\Big)\nonumber
   \end{align}
 
\begin{align}
%%%%%%%%%%%%%%%%%%%%%%%%%%%%%%%%%%%%%%%%%%%%%%%%%%%%%%%%%%
          &+ 
       12 \big (6\, m_{q_ {13}}\, m_{q_ {23}} + 9\, m_{q_ {13}}\, m_c + 9\, m_{q_ {23}}\, m_c - 
          8\, m_{q_ {3}}\, m_c \big)\, I[-2, 2] + 9\, I[-2, 3]\bigg) - 
    27\, (m_{q_ {13}} + m_{q_ {23}})\, I[-1, 2]\Bigg)\nonumber\\
    &+ 32\, I[0, 3]
  \Bigg]\nonumber\\
   %%%%%%%%%%%%%%%%%%%%%%%%%%%%%%%%%%%%%%%%%%%%%%%%%%%%%%%%%%
&+\frac{e_{q_{13}}\,m_c\, \langle g_s^2G^2\rangle}{4718592\, \pi^8}
  \Bigg[  3\, (m_{q_ {13}} - m_{q_ {23}})\, m_c^8 \,I[-5, 2] + 
 2\, m_c^5\, \Big (6\, m_{q_ {13}}\, m_{q_ {2}} - 6\, m_{q_ {12}}\, m_{q_ {23}} - 5\, m_{q_ {12}}\, m_c + 
    5\, m_{q_ {2}}\, m_c \Big)\nonumber\\
    & \times I[-4, 2] + 
 12\, m_c^3\, \Big (\,(-\,2\, m_{q_ {2}} + m_{q_ {23}})\, m_c - m_{q_ {13}}\, (\,2\, m_{q_ {2}} + m_c) + 
    2\, m_{q_ {12}}\, (m_{q_ {23}} + m_c)\Big)\, I[-3, 2] \nonumber\\
    &+ 
 6 m_c \Big (-\,2 m_{q_ {12}}\, m_{q_ {23}} - 3\, m_{q_ {12}}\, m_c + 3\, m_{q_ {2}}\, m_c - 2\, m_{q_ {23}}\, m_c + 
    2\, m_{q_ {23}}\, (m_{q_ {2}}\, + m_c)\Big)\, I[-2, 2]+ \Big (4\, m_{q_ {12}}   \nonumber\\
    &- 3\, m_{q_ {13}}
    - 4 \,m_{q_ {2}} + 3 \,m_{q_ {23}}\Big)\, I[-1, 2] \Bigg]\nonumber\\
       %%%%%%%%%%%%%%%%%%%%%%%%%%%%%%%%%%%%%%%%%%%%%%%%%%%%%%%%%%
       %%%%%%%%%%%%%%%%%%%%%%%%%%%%%%%%%%%%%%%%
 &+\frac{e_{q_{23}}\,m_c\, \langle g_s^2G^2\rangle}{4718592\, \pi^8}
  \Bigg[  3\, (-m_{q_ {13}} + m_{q_ {23}})\, m_c^8\, I[-5, 2]  
 -2\, m_c^5\, \Big (6\, m_{q_ {12}}\, m_{q_ {13}} - 6\, m_{q_ {1}}\, m_{q_ {23}} - 5\, m_{q_ {1}}\, m_c + 
    5\, m_{q_ {12}}\, m_c\Big)\nonumber\\
    & \times I[-4, 2] + 
 12\, m_c^3 \,\Big ((m_{q_ {13}} - m_{q_ {23}})\, m_c + 2\, m_{q_ {12}}\, (m_{q_ {13}} + m_c) - 
    2\, m_{q_ {1}} \,(m_{q_ {23}} + m_c)\Big)\, I[-3, 2] \nonumber\\
    &-
 6\, m_c\, \Big (2\, m_{q_ {12}}\, m_{q_ {13}} - 2\, m_{q_ {1}}\, m_{q_ {23}} - 3 \,m_{q_ {1}}\, m_c 
 + 3\, m_{q_ {12}}\, m_c + 
    2\, m_{q_ {13}}\, m_c - 2\, m_{q_ {23}}\, m_c\Big)\, I[-2, 
   2] - \Big (4\, m_{q_ {1}} \nonumber\\
   &- 4\, m_{q_ {12}} - 3\, m_{q_ {13}} + 3\, m_{q_ {23}}\Big) \,I[-1, 2] \Bigg]\nonumber \\
%%%%%%%%%%%%%%%%%%%%%%%%%%%%%%%%%%%%
&+\frac{\,e_s\, \langle g_s^2G^2\rangle  }{21233664\, \pi^8} \Bigg[-4\,m_c^8\,\Bigg(
3\,\big( m_{q_ {13}} -  m_{q_ {23}}\big)^2\, I[-5, 2] + 
 32\, I[-5, 3] + 76\, f_{3\gamma}\, \pi^2\, I[-5, 2] \,\psi^a[u0]\Bigg)\nonumber\\
 %%%%%%%%%%%%%%%%%%%%%%%%
 &
 +4\,m_c^7\,\Bigg(
3\,\big( m_{q_ {13}} -  m_{q_ {23}}\big)^2\, I[-5, 2] + 
 3 \big(9\, m_{q_ {1}} - 18\, m_{q_ {12}} +9 \, m_{q_ {2}} - 64 m_{q_ {3}}\big)\, I[-4, 2] - 
 32\, f_{3\gamma}\,\pi^2\, \big(m_{q_ {13}} + m_{q_ {23}}\nonumber\\
 &- 4\, m_{q_ {3}}\big)\,  I[-4, 1]\,\psi^a[u_0]\Bigg)
 %%%%%%%%%%%%%%%%%%%%%%%%%%%%%%%%%%%%%%%%%
 + 8 \,m_c^6 \Bigg(
 9 \Big (8\, m_{q_ {12}}^2 + 8\, m_{q_ {1}}\,m_{q_ {2}} + \big (m_{q_ {13}} - m_{q_ {23}}\big)^2\Big)\, I[-4, 
   2] + 2\, f_ {3\gamma} \,\pi^2 \Big (3\, (m_{q_ {13}} + m_{q_ {23}})^2\nonumber\\
   &\times I[-4, 1] - 20\, I[-4, 2]\Big)\, \psi^a[u_0] \Bigg)
   %%%%%%%%%%%%%%%%%%%%%%%%%%%%%%%%%%%%%%%%
   -3\,m_c^5  \Bigg(
   9\, \Big (3\, \big (m_{q_ {1}} - 2\, m_{q_ {12}} + m_{q_ {2}}\big) - 16 \, m_{q_ {3}}\Big)\, 
I[-3, 2] + 
 4\, f_ {3\gamma}\pi^2\, \Big (9\, m_{q_ {1}} \nonumber\\
 &+ 18\, m_{q_ {12}} - 24\, m_{q_ {13}} + 9\, 
  m_{q_ {2}} -24\, m_{q_ {23}} + 64\, m_{q_ {3}}\Big)\,  I[-3, 1]\, \psi^a[u_0]\Bigg)
  %%%%%%%%%%%%%%%%%%%%%%%%%%%%%%%%%%%%%%%%%%%%%%%%%%%%%%
   -m_c^4 \Bigg(
   27 \big (32\, m_{q_ {12}}^2 + 32\, m_{q_ {1}}\, 
   m_{q_ {2}} + 3\, (m_{q_ {13}} \nonumber\\
   &- m_{q_ {23}})^2\big)\, I[-3, 2] - 128 \, 
I[-3, 3] + 
 36\,  f_ {3\gamma}\, \pi^2\, \Big (\, 
    4\, \big (8\, m_{q_ {12}}^2 + 8 \,m_{q_ {1}} \,m_{q_ {2}} - (m_{q_ {13}} + m_{q_ {23}})^2\big) I[-3, 
       1] + I[-3, 2]\Big)\nonumber\\
       &\times \psi^a[u_0]   \Bigg)
       %%%%%%%%%%%%%%%555555555555555
  +3\, m_c^3 \Bigg(
   \, \big(\,27\, (m_{q_ {1}} - 2\, m_{q_ {12}} + m_{q_ {2}}) - 128\, m_{q_ {3}}\big) \, I[-2, 2] + 
 24\, f_ {3\gamma} \Big (\, 
   3\, \big (m_{q_ {1}} + 2\, m_{q_ {12}} - 4\, m_{q_ {13}}\, +m_{q_ {2}} \nonumber\\
   &- 4\, m_{q_ {23}}\big) + 
    16 \, \pi^2\, m_{q_ {3}}\Big) \,I[-2, 1] \, \psi^a[u_0]\Bigg)
    %%%%%%%%%%%%%%%%%%%%%%%%%%%%%%%%%%%%%%
    +36\,m_c^2\, \Bigg(
    \Big (12\, m_{q_ {12}}^2 + 12 \,m_{q_ {1}}\, m_{q_ {2}} + (m_{q_ {13}} - m_{q_ {23}})^2\Big)\,I[-2, 2] \nonumber\\
    &+  2 \, I l[-2, 3] + 
 f_ {3\gamma} \pi^2 \Big (\big (-32\, m_{q_ {12}}^2 - 32\, m_{q_ {1}}\, m_{q_ {2}} + 
        3 \,(m_{q_ {13}} + m_{q_ {23}})^2\big)\, I[-2, 1] - 8 \, I[-2, 2]\Big) \psi^a[u_0]
    \Bigg)\nonumber\\
    %%%%%%%%%%%%%%%%%%%%%%%%%%%%%%%%%55
  &  -192\, m_c\, \Bigg(f_{3\gamma} \big(m_{q_ {13}} + m_{q_ {23}}\big)\, I[-1, 1] \,\psi^a[u_0] \Bigg)
  %%%%%%%%%%%%%%%%%%%%%%%%%%%%%%%%5
  +192 \Bigg(f_{3\gamma} \Big(-5 (m_{q_ {13}} + m_{q_ {23}}) + 4 m_{q_ {3}}\Big)\,I[0, 1]\, \psi^a[u0]\Bigg)\Bigg]\nonumber\\
     %%%%%%%%%%%%%%%%%%%%%%55
     &+\frac{e_s \, \langle \bar q_1 q_1 \rangle}{18432\, m_c^2\, \pi^6}\Bigg[
      -6\, m_{q_{2}}\, m_c^5 \Big (2\, m_c^5\, I[-5, 3] + 6\, m_{q_{3}}\, m_c^4 \,I[-4, 2] + 
    m_c^3\, \big(3\, m_0^2\, I[-4, 2] + 4\, I[-4, 3]\big) \nonumber\\
    &- 
    6 \,m_{q_{3}}\, m_c^2\, \big(m_0^2\, I[-3, 1] + 2\, I[-3, 2]\big) + 
    m_c\, \big(-3\, m_0^2\, I[-3, 2] + 2\, I[-3, 3]\big) + 
    6\, m_{q_{3}}\, \big(I[-2, 2]-m_0^2\, I[-2, 1] \big)\Big) \nonumber
            \end{align}

\begin{align}
&- 48\, m_{q_{2}}\, I[0, 3] +
  m_{q_{1}} \Bigg (\, m_c^5\, \Big (-2\, m_c^7\, I[-6, 3] + 3\, m_0^2\, m_c^5 \,I[-5, 2] + 
        6 \,m_{q_{3}}\, m_c^6 \,I[-5, 2] + 6\, m_0^2\, m_{q_{3}}\, m_c^4\,I[-4, 1] \nonumber\\
& + 
        6\, m_c^3\, I[-4, 3] - 3\, m_0^2\, m_c\, I[-3, 2] - 
        18\, m_{q_{3}}\, m_c^2\, I[-3, 2] + 4\, m_c \,I[-3, 3]- 
        6\, m_0^2\, m_{q_{3}}\,I[-2, 1] + 12\, m_{q_{3}}\, I[-2, 2]\Big) \nonumber\\
        &  +  24\, m_0^2\, m_{q_{3}}\, m_c\, I[0, 1] + 8\, I[0, 3]\Bigg) +
     4\, f_ {3\gamma}\, \pi^2 \Bigg (m_c^5\bigg (m_c \Big (m_{q_{1}} \big(3\, m_c^4 \,
     I[-5, 2] + 2\, m_c^2 \big(-m_0^2 + 3\, m_{q_{3}} m_c\big)\, I[-4, 1] \nonumber\\
&-              3\, I[-3, 2]\big) - 
          6 \,m_{q_{2}}\, \big(2\, m_c^2\, I[-4, 2] + m_0^2\, I[-3, 1] - 
              4\, m_{q_{3}}\, m_c \,I[-3, 1]- 2\, I[-3, 2]\big)\Big) - 
       6\, \big(m_{q_{1}} - 4 m_{q_{2}}\big)\, m_{q_{3}}   \nonumber\\
              &\times I[-2, 1]\bigg)- 
    2\, m_0^2\, \big(m_{q_{1}} - 3\, m_{q_{2}}\big) \,m_{q_{3}}\, m_c\, I[0, 0] + 
    2 \,\big(m_0^2 (m_{q_{1}} - 3 m_{q_{2}}) + 12\, m_{q_{1}}\, m_{q_{3}}\, m_c\big)\, I[0, 
       1]\Bigg) \psi^a[u_0]
     \Bigg]    \nonumber\\
     %%%%%%%%%%%%%%%%%%%%%%%%%%%%%%%%%%%%%%%%%%%
     %%%%%%%%%%%%%%%%%%%%%%%%%%%%%%%%%%%%%%%%%%%
      %%%%%%%%%%%%%%%%%%%%%%%%%%%%%%%%%%%%%%%%%%%%
&-\frac{e_s \,m_{q_{12}}\, \langle \bar q_{12} q_{12} \rangle}{9216\, m_{c}^2\, \pi^6}\Bigg[
     2\, m_c^{12}\, I[-6, 3] - 6\, m_{q_{3}}\, m_c^{11}\, I[-5, 2] - 
 3\, m_c^{10}\, (m_0^2\, I[-5, 2] - 4\, I[-5, 3]) - 
 6\, m_{q_{3}}\, m_c^9 \big(m_0^2 \nonumber\\
 & \times I[-4, 1] - 6\, I[-4, 2]\big) + 
 18\, m_c^8\, \big(m_0^2\, I[-4, 2] + I[-4, 3]\big) - 
 18\, m_{q_{3}}\, m_c^7 \,\big(2\, m_0^2 I[-3, 1] + 3\, I[-3, 2]\big) \nonumber\\
 &+ 
 m_c^6\, \big(-15\, m_0^2\, I[-3, 2] + 8\, I[-3, 3]\big) - 
 6\, m_{q_{3}}\, m_c^5 \,\big(5\, m_0^2\, I[-2, 1] - 4\, I[-2, 2]\big) - 
 24\, m_0^2\, m_{q_{3}}\, m_c\, I[0, 1]\nonumber\\
 &+ 40\, I[0, 3] - 
 4\, f_{3\gamma}\, \pi^2\, \Bigg (3\, m_c^{10}\, I[-5, 2] + 
     6\, m_{q_{3}}\, m_c^9\, I[-4, 1] - 
     2\, m_c^8\, \big(m_0^2\, I[-4, 1] + 6\, I[-4, 2]\big) + 
     24\, m_{q_{3}}\, m_c^7\nonumber\\
     & \times I[-3, 1] + 
     m_c^6\, \big(-6\, m_0^2 \,I[-3, 1] + 9\, I[-3, 2]\big) + 
     18\, m_{q_{3}}\, m_c^5\, I[-2, 1] - 4 \,m_0^2 \,I[0, 1] + 
     4\, m_{q_{3}}\, m_c\, \big(m_0^2\, I[0, 0] \nonumber\\
     &+ 6\, I[0, 1]\big)\Bigg) \psi^a[u_ 0]
     \Bigg]\nonumber\\
%%%%%%%%%%%%%%%%%%%%%%%%%%%%%%%%%
&+\frac{e_s \, \langle \bar q_{3} q_{3} \rangle}{9216\, m_{c}^2\, \pi^6}\Bigg[
m_c^5 \Bigg (m_c^8 \,I[-6, 3] + 3\, m_0^2\, m_c^6 \,I[-5, 2] - 
    6\, m_c^4\, \big(\,3\, (m_{q_{12}}^2 + m_{q_{1}}\, m_{q_{2}})\, I[-4, 2] + I[-4, 3]\big)\nonumber\\
    &+     m_c^2 \Big (18\, m_0^2\, \big(m_{q_{12}}^2 + m_{q_{1}}\, m_{q_{2}}\big)\, I[-3, 1] - 
       9\, \big(m_0^2 - 4\, (m_{q_{12}}^2 + m_{q_{1}}\, m_{q_{2}})\big)\, I[-3, 2] - 
       8 I[-3, 3]\Big) \nonumber\\
       &+ 18\, m_0^2\, \big(m_{q_{12}}^2 + m_{q_{1}}\, m_{q_{2}}\big)\, I[-2, 1] + 
    6 \,\big (m_0^2 - 3 \,(m_{q_{12}}^2 + m_{q_{1}}\, m_{q_{2}})\big)\, I[-2, 2] - 
    3 \,I[-2, 3]\,\Bigg) \nonumber\\
    &- 
 m_{q_{3}} \Bigg (m_c^6 \bigg(2\, m_c^6\, I[-6, 3] - 3\, m_0^2\, m_c^4\, I[-5, 2] + 
       6\, m_c^2\, \Big(\,3 \,\big(m_{q_{12}}^2 + m_{q_{1}}\, m_{q_{2}}\big)\, I[-4, 2] 
       - I[-4, 3]\Big) \nonumber\\
       &+ 
       6\, m_0^2\, \big(m_{q_{12}}^2 + m_{q_{1}}\, m_{q_{2}}\big)\, I[-3, 1] + 
       3\, \big(m_0^2 - 6 \,(m_{q_{12}}^2 + m_{q_{1}}\, m_{q_{2}})\big)\, I[-3, 2] 
       - 4\, I[-3, 3]\bigg) + 
    6\, m_0^2\, \big(m_{q_{12}}^2 \nonumber\\
    &+ m_{q_{1}}\, m_{q_{2}}\big) \,I[0, 1] - 8\, I[0, 3]\Bigg) - 
 16\, m_c\, I[0, 3] + 
 4\, f_{3\gamma}\, \pi^2\, \Bigg (\,m_c^5 \,\bigg (m_c^2\, \Big(2\, m_c^4\, I[-5, 2] + 
           3\, m_0^2\, m_c^2\, I[-4, 1] \nonumber\\
           &+ 
           12 \big(m_{q_{12}}^2 + m_{q_{1}}\, m_{q_{2}}\big) \,I[-3, 1] - 6\, I[-3, 2]\Big) + 
        m_{q_{3}}\, m_c \Big (3\, m_c^4\, I[-5, 2] - 
           2\, m_0^2\, m_c^2 \,I[-4, 1] - 
           6\, \big(m_{q_{12}}^2 \nonumber\\
           &+ m_{q_{1}}\, m_{q_{2}}\big)\, I[-3, 1] - 3\, I[-3, 2]\Big) - 
        3\, \Big(m_0^2 - 4\, (m_{q_{12}}^2 + m_{q_{1}}\, m_{q_{2}})\Big) \,I[-2, 1] + 
        4\, I[-2, 2]\bigg) + 
     3\, m_0^2 \,\big(m_{q_{12}}^2 \nonumber\\
     &+ m_{q_{1}}\, m_{q_{2}}\big) \,m_c\, I[0, 0] + 
     2 \,\Big(-3\,\big(m_{q_{12}}^2 + m_{q_{1}}\, m_{q_{2}}\big)\, m_{q_{3}} + m_0^2\, \big(m_{q_{3}} 
     + 6 m_c\big)\Big) I[0, 
        1]\Bigg) \psi^a[u_ 0]
     \Bigg]
    \Bigg\},\\ 
    \nonumber\\
 F_2&=\frac{m_{P_{c \bar s}}\,e^{m^2_{P_{c \bar s}}/M^2}}{\lambda_{P_{c \bar s}}^2}\Bigg\{
 \frac{e_s\,\langle \bar ss \rangle}{5898240 \,m_c^2\, \pi^6}\Bigg[
 160\, \Bigg (-m_c^5 \bigg (m_c^7\, I[-6, 4] - 
       4\, m_{q_{3}}\, m_c^6 \, I[-5, 3] + 
       3\, m_c^3 \Big (4\, \big(m_{q_{12}}^2 + m_{q_{1}}\, m_{q_{2}}\big)\nonumber\\
       & \times I[-4, 3] - 
          I[-4, 4]\Big) + 
       12\, m_{q_{3}}\, m_c^2 \Big (-3\, \big(m_{q_{12}}^2 + m_{q_{1}}\, m_{q_{2}}\big)\, I[-3, 2] + 
          I[-3, 3]\Big) + 
       2\, m_c \big (\,6\, (m_{q_{12}}^2 + m_{q_{1}}\, m_{q_{2}})\, I[-3,3]\nonumber\\
       &  + 
          I[-3, 4]\big) + 36\,m_{q_{3}}\, \big(m_{q_{12}}^2 + m_{q_{1}} m_{q_{2}}\big)\,  I[-2, 2] + 
       8\, m_{q_{3}}\, I[-2, 3]\bigg) - 
    8\,\big (3\, m_{q_{12}}^2 + 3\, m_{q_{1}}\, m_{q_{2}} + 2\, m_{q_{3}}\, m_c\big) I[0, 3]\Bigg)A[u_ 0]\nonumber\\
    &  + 
 32 \,\chi\, \Bigg (m_c^{14} \,I[-7, 5] + 5\, m_{q_{3}}\, m_c^{13}\, I[-6, 4] + 
     8\, m_c^{12}\, I[-6, 5] - 40\, m_{q_{3}}\, m_c^{11}\, I[-5, 4] + 
     6\, m_c^{10} \Big (\,5 \,\big(m_{q_{12}}^2+ m_{q_{1}}\, m_{q_{2}}\big) \nonumber
       \end{align}
       \begin{align}
       & \times I[-5, 4]+ 
        3 \,I[-5, 5]\Big) + 
     30\, m_{q_{3}}\, m_c^9 \Big (\,4 \,\big(m_{q_{12}}^2 + m_{q_{1}}\, m_{q_{2}}\big)\, I[-4, 3] + 
        3\,I[-4, 4]\Big) + 
     4\, m_c^8 \Big (-15\, \big(m_{q_{12}}^2 \nonumber\\
     & + m_{q_{1}} m_{q_{2}}\big)\, I[-4, 4] + 
        4\, I[-4, 5]\Big) + 
     80\, m_{q_{3}}\, m_c^7 \Big (\,3\, \big(m_{q_{12}}^2 + m_{q_{1}}\, m_{q_{2}}\big) \,I[-3, 3] - 
        I[-3, 4]\Big) + 
     5\, m_c^6 \Big (\,6\, \big(m_{q_{12}}^2 \nonumber\\
     &+ m_{q_{1}}\, m_{q_{2}}\big)\, I[-3, 4] + 
        I[-3, 5]\Big) + 
     5 \,m_{q_{3}}\, m_c^5 \Big (24 \big(m_{q_{12}}^2 + m_{q_{1}} \,m_{q_{2}}\big)\, I[-2, 3] + 
        5\, I[-2, 4]\Big) + 
     480\, \big(m_{q_{12}}^2+ m_{q_{1}}\, m_{q_{2}}\big) \nonumber\\
     & \times m_{q_{3}}\, m_c\, I[0, 3] + 
     48\, I[0, 5]\Bigg) \varphi_\gamma[u_ 0] + 
 5\, \bigg (7\, m_c^{12}\, I[-6, 4] + 
    2\, \big (\,9 \,m_{q_{1}} + 18\, m_{q_{12}} + 6\, m_{q_{13}} + 9\, m_{q_{2}} + 6\, m_{q_{23}}- 
       32\, m_{q_{3}}\big) \nonumber\\
    &\times m_c^{11}\, I[-5, 3] + 
    6\, m_c^{10}\, \Big (\big (-m_{q_{13}}^2 + 10\, m_{q_{13}}\,m_{q_{2}} + 10\, m_{q_{1}}\, m_{q_{23}} - 
          2\, m_{q_{13}}\, m_{q_{23}} - m_{q_{23}}^2+ 10\, m_{q_{12}}\, (m_{q_{13}}+ m_{q_{23}})\big)  \nonumber\\
          %%%%%%%%%%%%%%%%%%%%%%%%%%%%%%%%%%%%%%%%%%%%%%%%%%%%%%%%%%%%%%%%%%%%
       &\times I[-5,3] + 4 \,I[-5, 4]\Big)+ 
    12\,m_c^9\, \big (3\, m_{q_{1}} + 6\, m_{q_{12}} + 8\, m_{q_{13}} + 3\, m_{q_{2}} + 
       8 \,m_{q_{23}}\big)  \,I[-4, 3] - 84\, m_c^8\,I[-4, 4] + 
    12 m_c^7 \Big (\nonumber\\
       &  9\, \big (m_{q_{13}}^2 m_{q_{2}} + 2 \,m_{q_{12}}\, m_{q_{13}}\, m_{q_{23}} + 
          m_{q_{1}}\, m_{q_{23}}^2\big)\, I[-3, 2] + \big (\, 
          3\, (m_{q_{1}} + 2\, m_{q_{12}} + 6\, m_{q_{13}} + m_{q_{2}}+ 6 \,m_{q_{23}}) + 
           16 \,m_{q_{3}}\big) \nonumber\\
          &\times I[-3, 3]\Big)+ 
    2\, m_c^6 \Big (-\, 
      9 \, \big (m_{q_{13}}^2 + 6\, m_{q_{13}}\, m_{q_{2}} + 6 \,m_{q_{1}} \,m_{q_{23}} + 2\, m_{q_{13}}\, m_{q_{23}} + 
          m_{q_{23}}^2 + 6\, m_{q_{12}}\, (m_{q_{13}}+ m_{q_{23}})\big) I[-3, 3] \nonumber\\
          & + 
       34\, I[-3, 4]\Big)+ 
    4\, m_c^5 \Big (-54 \,\big (m_{q_{13}}^2 \,m_{q_{2}} + 2\, m_{q_{12}}\, m_{q_{13}}\, m_{q_{23}} + 
          m_{q_{1}}\, m_{q_{23}}^2\big)\, I[-2,2] + \big (9\, m_{q_{1}} 
         + 18\, m_{q_{12}} + 48\, m_{q_{13}}  \nonumber\\
         &+ 9\, m_{q_{2}}+ 48\, m_{q_{23}} + 
          32\, m_{q_{3}}\big)\, I[-2, 3]\Big) - 
    3\, m_c^4 \Big (\,8\, \Big (\,m_{q_{13}} \big(2\, m_{q_{12}} + m_{q_{13}} + 2\, m_{q_{2}}\big) + 
          2 \,\big(m_{q_{1}} + m_{q_{12}} + m_{q_{13}}\big)\, m_{q_{23}} \nonumber\\
             &+ m_{q_{23}}^2\Big)\, I[-2, 3] + 
       5\, I[-2, 4]\Big) + 
    6 \,m_c^3 \bigg (\,18\, \Big (m_{q_{13}}^2\, m_{q_{2}} + 2\, m_{q_{12}}\, m_{q_{13}} \,m_{q_{23}} + 
          m_{q_{1}}\, m_{q_{23}}^2\Big)\, I[-1, 2] 
          + \Big (3\, m_{q_{1}} + 6 \,m_{q_{12}} \nonumber\\
          & + 10\, m_{q_{13}} + 3\, m_{q_{2}} + 
          10\, m_{q_{23}}\Big) \,I[-1, 3]\bigg) - 
    48 \bigg (m_{q_{13}} \big(2 m_{q_{12}} + m_{q_{13}} + 2 m_{q_{2}}\big)+ 
       2\, \Big(m_{q_{1}} + m_{q_{12}}+ m_{q_{13}}\Big) \,m_{q_{23}}+ m_{q_{23}}^2\bigg)  \nonumber\\
    &\times  I[0, 3] + 
    16 \big (9 (m_{q_{1}} + 2 m_{q_{12}} + 4 m_{q_{13}} + m_{q_{2}} + 4 m_{q_{23}}) + 
       16 m_{q_{3}}\big) m_c\, I[0, 3]\Bigg)\, I_1[\tilde S] \nonumber\\
       %%%%%%%%%%%%%%%%%%%%%%%%%%%%%%
       %%%%%%%%%%%%%%%%%%%%%%%%%%%%%%%%%%%%%%%%%
      & + 
 80 \Bigg (\,m_c^5\, \bigg (-m_c^9\, I[-7, 4] - 
       4 \,m_{q_{3}}\, m_c^8 \,I[-6, 3] + 
       6\, m_c^5 \Big (\,4\, \big(m_{q_{12}}^2 + m_{q_{1}}\, m_{q_{2}}\big) \,I[-5, 3] + 
          I[-5, 4]\Big) + 
       24\, m_{q_{3}}\, m_c^4 \Big (3 \big(m_{q_{12}}^2 \nonumber\\
       &+ m_{q_{1}}\, m_{q_{2}}\big)\, I[-4, 2] + 
          I[-4, 3]\Big) + 
       8\, m_c^3 \Big (\,6 \,\big(m_{q_{12}}^2 + m_{q_{1}}\, m_{q_{2}}\big)\, I[-4, 3] - 
          I[-4, 4]\Big) + 
       16\, m_{q_{3}}\, m_c^2 \Big (-9 \,\big(m_{q_{12}}^2 + m_{q_{13}}^2\nonumber\\
       &+ m_{q_{23}}^2\big) \Big)\, I[-3, 2] +  2\, I[-3, 3] + 
       3\, m_c \Big (\,8\, \big(m_{q_{12}}^2 + m_{q_{1}}\, m_{q_{2}}\big)\, I[-3, 3] + 
          I[-3, 4]\Big) + 
       12\, m_{q_{3}} \Big (\,6\, \big(m_{q_{12}}^2 + m_{q_{1}} m_{q_{2}}\big)\, I[-2, 2] \nonumber\\
       &+ I[-2, 3]\Big)\bigg) + 
    32 \Big (\,3\, m_{q_{12}}^2 + 3\, m_{q_{1}}\, m_{q_{2}} + 2\, m_{q_{3}}\, m_c\Big) \,I[0, 
       3]\Bigg)\, I_2[h_\gamma]
 \Bigg]
  \Bigg\},
\end{align}
and
\begin{align}
 F_3&=\frac{4\,m_{P_{c \bar s}}\,e^{m^2_{P_{c \bar s}}/M^2}}{\lambda_{P_{c \bar s}}^2}\Bigg\{ 
 \frac{e_s \, \langle \bar ss \rangle}{73728\, m_c^2\, \pi^6}\Bigg[
 4\, A[u_ 0] \Bigg (m_c^5 \bigg (m_c \Big (m_c^6\, I[-6, 3] - 
          3\, m_ {q_ {3}} \,m_c^5 \, I[-5, 2] + 3\, m_c^4 \,I[-5, 3] + 
          9\, m_ {q_ {3}}\, m_c^3\nonumber\\
          &\times I[-4, 2] + 3\, m_c^2\, I[-4, 3] - 
          9\, m_ {q_ {3}}\, m_c I[-3, 2] + I[-3, 3]\Big) + 
       3\, m_ {q_ {3}}\, I[-2, 2]\bigg) + 
    8\, I[0, 3]\Bigg) + \chi\, m_c\, \Bigg (m_c^4 \bigg (m_c \Big (m_c^8 \nonumber\\
    & \times I[-7, 4] 
    + 4\, m_ {q_ {3}}\, m_c^7\, I[-6, 3] - 4\, m_c^6\, I[-6, 4] + 
           16\, m_ {q_ {3}}\, m_c^5\, I[-5, 3] + 6 \,m_c^4 I[-5, 4] + 
           24\, m_ {q_ {3}}\, m_c^3 \,I[-4, 3]- 4\, m_c^2  \nonumber\\
           &\times I[-4, 4] + 
           16\, m_ {q_ {3}}\, m_c \,I[-3, 3] + I[-3, 4]\Big) + 
        4\, m_ {q_ {3}}\, I[-2, 3]\bigg) + 
     64\, m_ {q_ {3}}\, I[0, 3]\Bigg) \varphi_\gamma[u_ 0] - 
 2 \Bigg (m_c^5 \bigg (m_c \Big (m_c^8\, I[-7, 3]\nonumber\\
 & + 
          3\, m_ {q_ {3}}\, m_c^7 \, I[-6, 2] + 4 \,m_c^6 \, I[-6, 3] - 
          12 \, m_ {q_ {3}}\, m_c^5 \, I[-5, 2] + 6\, m_c^4\, I[-5, 3] + 
          18 \, m_ {q_ {3}}\, m_c^3 \,I[-4, 2] + 4\, m_c^2 \,I[-4, 3]  \nonumber\\
        &-  12\, m_ {q_ {3}}\, m_c \, I[-3, 2] + I[-3, 3]\Big) + 
       3 \,m_ {q_ {3}} \, I[-2, 2]\bigg) + 16 \,I[0, 3]\Bigg) I_2[h_\gamma]\Bigg]
 \Bigg\},
 \end{align}
where $ s_0 $ is the continuum threshold, $u_0= \frac{M_1^2}{M_1^2+M_2^2} $, $ \frac{1}{M^2}= \frac{1}{M_1^2}+\frac{1}{M_2^2}$ with $ M_1^2 $ and $ M_2^2 $ being the Borel parameters in the initial and final states, respectively  and we have not presented the explicit form of $F_4$ as it gives contributions only to the magnetic octupole moment, 
whose value is roughly zero.
Here $e_q$ is the electric charge of the corresponding quark; and 
%$\chi$ is the magnetic susceptibility of the quark condensate,
%$m_0^2 =\langle \bar q\, g_s\, \sigma_{\alpha\beta}\, G^{\alpha\beta}\, q \rangle /\langle \bar qq \rangle $, 
$\langle \bar qq \rangle$ and $\langle g_s^2 G^2\rangle$ are quark and gluon condensates, respectively. We should also remark that, in the above sum rules, for simplicity we have only presented the terms that give considerable contributions to the numerical values of the quantities under consideration and ignored to present many higher dimensional operators although they have been considered in the numerical analyses. In the presented results terms with gluon condensate multiply high twist (twist-3) DAs of photon come from the nonperturbative contributions in QCD side. Such that one of the quarks interact with the photon nonperturbatively and two single gluon fields from two different propagators make gluon condensate and the remaining two propagators are replaced by their free parts.
The values of $e_{q_i}$, $e_{q_{ij}}$,  $m_{q_i}$ and  $m_{q_{ij}}$ 
corresponding to different states are given in Table II.
 \begin{table}[htp]
	\addtolength{\tabcolsep}{8pt}
	      \begin{tabular}{c|c|c|c|c|c|c|c|c|c|c|c|cccccc}
            \hline\hline
            ${P_{c\bar s}}$          & $e_{q_1}$  & $e_{q_2}$ & $e_{q_3}$&$e_{q_{12}}$&  $e_{q_{13}}$&  $e_{q_{23}}$&  
            $m_{q_1}$ & $m_{q_2}$ & $m_{q_3}$ & $m_{q_{12}}$ & $m_{q_{13}}$ & $m_{q_{23}}$ \\
            \hline\hline
	    $uuuc\bar s$     & $e_u$      &$e_u$      &  $e_u$   &$e_u$&$e_u$ &$e_u$&0&0&0&0&0&0 \\
	    $dddc\bar s$     & $e_d$      &$e_d$      &  $e_d$   &$e_d$&$e_d$ &$e_d$&0&0&0&0&0&0\\
	    $sssc\bar s$     & $e_s$      &$e_s$      &  $e_s$   &$e_s$&$e_s$  &$e_s$&$m_s$&$m_s$&$m_s$&$m_s$&$m_s$&$m_s$\\
	    $uddc\bar s$     & $e_u$      &$e_d$      &  $e_d$   &0&$0$&$e_d$&0&0&0&0&0&0\\
	    $ussc\bar s$     & $e_u$      &$e_s$      &  $e_s$   &0&$0$&$e_s$&0&0&0&0&0&$m_s$\\
	    $duuc\bar s$     & $e_d$      &$e_u$      &  $e_u$   &0 &0&$e_u$&0&0&0&0&0&0\\
	    $dssc\bar s$     & $e_d$      &$e_s$      &  $e_s$   &0 &0&$e_s$&0&0&0&0&0&$m_s$\\
	    $suuc\bar s$     & $e_s$      &$e_u$      &  $e_u$   &0 &0&$e_u$&0&0&0&0&0&0\\
	    $sddc\bar s$     & $e_s$      &$e_d$      &  $e_d$   &0 &0&$e_d$&0&0&0&0&0&0\\
	    $udsc\bar s$     & $e_u$      &$e_d$      &  $e_s$   &0 &0&0&0&0&0&0&0&0\\
	    \hline\hline
\end{tabular}
\caption{The values of $e_{q_i}$, $e_{q_{ij}}$,  $m_{q_i}$ and  $m_{q_{ij}}$ 
related to the expressions of the electromagnetic form factors in Eqs.(17), (18) and (19).}
	\label{table2}
\end{table}
           
The functions~$I[n,m]$,~$I_1[\mathcal{A}]$ and~$I_2[\mathcal{A}]$  are
defined as:
\begin{align}
I[n,m]&= \int_{m_c^2}^{s_0} ds \int_{m_c^2}^s dl~ e^{-s/M^2}~\frac{(s-l)^m}{l^n},\nonumber\\
I_1[\mathcal{A}]&=\int D_{\alpha_i} \int_0^1 dv~ \mathcal{A}(\alpha_{\bar q},\alpha_q,\alpha_g)
 \delta(\alpha_{\bar q}+ v \alpha_g-u_0),\nonumber\\
I_2[\mathcal{A}]&=\int_0^1 du~ A(u) \nonumber.
 \end{align}

\bibliography{refs}
\end{document}